\documentclass[11pt]{cernrep}
\usepackage{graphicx}
\usepackage{epsf}
\usepackage{epsfig}

\usepackage[english]{babel}
\usepackage{amsmath}
\usepackage{amssymb}
\usepackage{units}
\usepackage{graphicx}
\usepackage{cite}
\usepackage{psfrag}
\usepackage{rotate}
\usepackage{amssymb}

\renewcommand{\thesection}{\arabic{section}}
\renewcommand{\thesubsection}{\arabic{section}.\arabic{subsection}}

\newcommand{\lsim}{
\mathrel{\hbox{\rlap{\hbox{\lower4pt\hbox{$\sim$}}}\hbox{$<$}}}}

\newcommand{\gsim}{
\mathrel{\hbox{\rlap{\hbox{\lower4pt\hbox{$\sim$}}}\hbox{$>$}}}}

\begin{document}
 
\pagestyle{empty}

%%%%%%%%%%% CERN Titlepage %%%%%%%%%%%

\begin{titlepage}
\begin{flushright}
\begin{tabular}{r}
CERN-PH-TH/2009-026\\
% hep-ph/?????????\\
\end{tabular}
\end{flushright}

\vspace*{1.9truecm}

\begin{center}
\boldmath
{\LARGE \bf Jet Quenching in Heavy Ion Collisions}
\unboldmath

\vspace*{2.6cm}

\smallskip
\begin{center}
{\sc {\large Urs Achim Wiedemann}}\\
\vspace*{2mm}
{\sl {\large Theory Unit, Department of Physics, CERN, 
CH-1211 Geneva 23, Switzerland}}
\end{center}

\vspace{2.6truecm}

\vspace{2.6cm}
 
{\sl 
Review article prepared for the Landolt-B\"ornstein volume on Relativistic Heavy Ion Physics
}

{\sl 
Editor: Reinhard Stock
}
\end{center}

\end{titlepage}
 
\thispagestyle{empty}
\vbox{}
\newpage

%%% end CERN title page %%%%%%%%%%%%%

\pagestyle{plain}

\setcounter{page}{1}
\pagenumbering{roman}

\tableofcontents

\newpage

\setcounter{page}{1}
\pagenumbering{arabic}

\title{Jet Quenching in Heavy Ion Collisions}
% \author{Urs Achim Wiedemann}
% \institute{CERN, Geneva, Switzerland}
% \maketitle
% \begin{abstract}
% We review ...
% \end{abstract}

\pagestyle{plain}

\section{INTRODUCTION}\label{sec:intro}
\setcounter{equation}{0}
Understanding the passage of particles through matter amounts to understanding
properties of the matter through which the particles pass. This is illustrated e.g. by the
Bethe-Bloch formula for the mean rate of energy loss $- dE/dx$. For electrically
charged particles,  $- dE/dx$ depends on several properties of matter, including its
atomic excitation and ionization properties and its polarizability, which can be
characterized by its plasma energy~\cite{Amsler:2008zzb}. 
In this way, the Bethe-Bloch formula provides, 
for instance, information about the suitability of solids, liquids and gases to serve as 
detector materials for characterizing the identity and energy of an impinging particle. 
Inverting this logic, 
the same medium-dependence could be used to characterize (unknown) properties 
of matter by measuring the energy loss of identified and well-calibrated particles. 
This inverse problem is of little practical relevance for the characterization of 
materials, whose properties are determined by the electrodynamical interaction,
and for whose characterization a broad array of other techniques exists. However, 
for the characterization of the strongly 
interacting, high-density matter produced in ultra-relativistic heavy ion collisions, 
alternative methods are scarce, and the study of medium-induced
parton energy loss and parton fragmentation has become one of the most promising
tools for a detailed characterization~\cite{Baier:2000mf,Kovner:2003zj,Gyulassy:2003mc,Jacobs:2004qv,CasalderreySolana:2007zz}. Although the tight interplay of theory and 
experiment has lead to rapid progress in recent years, the current understanding of 
medium-induced parton energy loss is incomplete, and any pure review is
likely to be outdated soon. As a consequence, the present article will focus mainly on 
the generic physics, which a complete theory of medium-induced parton energy loss 
should incorporate finally. The currently pursued heuristic approaches of modeling 
parton energy loss will be discussed only in this more general context. Their
comparison to data from RHIC is reviewed in another article in this volume~\cite{d'Enterria:2009am}.

Relativistic heavy ion collisions have been studied experimentally in the last decades
at increasing center-of-mass energies at the Brookhaven Alternating Gradient
Synchrotron AGS ($\sqrt{s_{_{NN}}} < 5$ GeV), the CERN Super Proton Synchrotron SPS
($\sqrt{s_{_{NN}}} \leq 20$ GeV) and the Brookhaven Relativistic Heavy Ion Collider RHIC
($\sqrt{s_{_{NN}}} \leq 200$ GeV). Soon, the Large Hadron Collider LHC at CERN will study 
heavy ion collisions at a center-of-mass energy $\sqrt{s_{_{NN}}} = 5.5$ TeV, which is almost 
a factor 30 higher than the maximal collision energy at RHIC. Here, we focus on collisions
at RHIC~\cite{Arsene:2004fa,Adcox:2004mh,Back:2004je,Adams:2005dq}
and at LHC~\cite{Carminati:2004fp,Alessandro:2006yt,D'Enterria:2007xr}
collider energies, which produce matter at the highest energy 
density in the largest volume and with the longest lifetime, which can be attained in any
laboratory experiment. More precisely, the initial volume is governed by the overlap of the 
colliding nuclei, the attained energy density exceeds the critical energy density for the phase 
transition to a Quark Gluon Plasma, and during its expansion the ultra-dense partonic system
spends at least the first few fm/c above this critical energy density. This is a long
lifetime compared to typical strong interaction times, and hence ultra-relativistic heavy ion 
collisions provide a unique opportunity for studying how properties of strongly interacting
non-abelian partonic matter emerge from the fundamental interactions of quantum
chromodynamics (QCD). But it is a lifetime far too short to test the produced system
with external probes. Thus, characterization of the properties of the produced system
must proceed by studying its decay products. At RHIC and even more so at 
LHC collider energies, partonic interactions at pertubatively large momentum 
transfer become abundant and the remnants of high energy partons 
% up to $O(10\, {\rm GeV})$ at RHIC and $O(100\, {\rm GeV})$ at the LHC 
become experimentally
accesssible. This gives access to a qualitatively novel class of  {\it auto-generated hard 
probes} of the produced matter. The idea behind this concept is that the identity and energy of
these initially produced parent partons can be calibrated independently, e.g. by the 
experimentally measured and pertubatively calculated yield of the hard process in proton-proton
collisions, or by the measured energy of the recoiling high-$p_T$ particle, or by 
calorimetric measurements which suitably subtract the high-multiplicity background of the heavy
ion collision. Hence, the partons produced in high-$Q^2$ processes in a heavy-ion collision
may be viewed as well-identified and calibrated projectiles, which suffer medium-induced 
energy loss while propagating through the dense matter produced in the collision. The
task is to characterize the medium-modification of the parton propagation and to relate it
to fundamental properties of the produced matter. Several aspects of this challenge 
are characteristically different from the problem of shooting an electrically charged particle
into normal matter. In particular:

\begin{enumerate}
\item {\it The projectile has a complicated time evolution even in the vacuum.}\\
In the standard Bethe-Bloch formula, the charged projectiles are asymptotic initial
states, which can be prepared in the infinite past. They are eigenstates of the QED
hamiltonian; so, in the absence of interactions with a medium, these particles do not
radiate or fragment. In contrast, the partons produced in high-$Q^2$ interactions are the 
seeds of final state parton showers; they are highly virtual and branch in the vacuum 
under the QCD time evolution into multi-parton final states. As a consequence, the
nature of the projectile changes as a function of the time passed since production.
What matters is not only whether the projectile interacts with a target, but
also when. 
\item {\it In parton energy loss problems, initial and final states have different physical degrees of freedom.}\\
For highly virtual partons, the initial stage of this
fragmentation process can be accounted for perturbatively. The non-perturbative latter
stage of jet fragmentation is hadronization. Fragmentation ends when hadronization is
complete, since hadrons are eigenstates of the QCD time evolution. The experimentally
accessible remnant of a parent parton, a.k.a. the jet, is a multi-hadron final state, and
its medium-modification must be established on top of this hadronic
structure. This is very different from QED, where elementary, electrically charged projectiles
such as electrons or muons do not change their identity during the interaction and 
the medium-modification of the projectile is fully 
characterized by $-dE/dx$ and by its final transverse momentum.
\item{\it The target is a source of momentum transfer and of color transfer, and this color must be 
bleached.}\\
Hadronization of a final state parton shower is a dynamical color neutralization process,
which bleaches colored partons into color-singlet hadrons. While a thorough dynamical
understanding of hadronization is lacking, we know that this process proceeds
locally in phase space. Hence, coarse features of the phase space distribution of jets
can be expected to be unmodified by hadronization. 
But QCD is a finite resolution theory, and for sufficiently fine
resolution, hadronization effects matter. Hence, while some characteristics of a jet may 
be sensitive solely to medium modifications which originate in the early partonic stage
of the heavy ion collision, other characteristics may be sensitive to the later
stage. Any color transfer bertween projectile and medium will affect the dynamics of
hadronization and thus jet hadrochemistry. 
\item{\it The target evolves strongly.}\\
From the combined analysis of soft single-inclusive hadron spectra, two-particle correlations
and their azimuthal dependence, one knows that the matter produced in a heavy ion 
collisions expands rapidly. Thus, the highly energetic partonic projectile propagates through
an environment whose energy density decreases rapidly. As we shall review below, however, 
parton energy loss can depend quadratically on the in-medium path length. In this case, 
interactions at late times are more efficient in degrading the energy of the leading
partonic projectile, and in some model scenarios this can compensate completely the
decreasing energy density of the expanding system. In any case, control over the
geometrical extension and dynamical expansion of the produced matter is important
for any quantitative comparison with data. 
\end{enumerate}

Up until recently, most experimental and theoretical work on parton energy loss 
focussed on single-inclusive hadron spectra at high transverse momenta 
($p_T > 5 -10\, {\rm GeV}$). Such hadrons can be regarded as the leading fragments
of parent partons and they are accompanied by a jet-like spray of subleading
particles. However, as discussed below, requiring detection of a high-$p_T$ 
hadron is a significant trigger bias on jet fragmentation, and the jet-like sprays
selected in this way differ strongly from samples of jets selected by calorimetric 
measurements. We expect that in comparison to high-$p_T$-triggered jet-like
multi-particle states, the internal structure of jets selected by 'true' calorimetric jet 
measurements is significantly more sensitive to properties of the medium, simply 
because any trigger bias of a partonic fragmentation pattern can obscure the imprints 
of a medium-dependence. For the purpose of this article, we draw from this the
radical consequence to discuss first qualitative expectations of the medium modification of true jets, 
before turning to the physics of single inclusive hadron spectra and their biases, and 
before discussing the current state of the art in the understanding and theoretical
modeling of jet quenching. 
%
%%%%%%%%%%%%%%%%%%%%%%%%%%%%%%%%%%%%%%%%%%%
\boldmath
\section{Jets in the absence and in the presence of a medium}
\label{sec2}
\setcounter{equation}{0}
\unboldmath
Hard processes are hadronic processes involving a large momentum transfer. Their
medium-dependence must be established on top of a reliable baseline. The obvious
baseline is the same hard process, measured in the absence of a medium, that is, in
elementary hadronic interactions such as proton-proton collisions. 
The study of hard processes in elementary collisions is one of the most successful 
textbook chapters of QCD, see e.g.~\cite{Mueller:1989hs,Ellis:1991qj,Sterman:1994ce}. 
Here, we recall in an eclectic manner only those few aspects, which are of particular 
relevance for the following discussion of medium effects. In particular, factorization
theorems ascertain that many hard processes can be factorized into the long distance 
physics of hadrons, which is not perturbatively calculable but processes-independent, 
and the short-distance
physics of partons, which is process-dependent but perturbatively calculable. For single-
inclusive hadron spectra in proton-proton collisions, for instance, this factorization takes the 
schematic form:
\begin{equation}
   d\sigma^{p+p\to h+X}
      =\sum_{f}
      d\sigma^{p+p\to f+X}
      \otimes 
    D_{f\to h}(z,\mu^2_F)\, ,
      \label{eq2.1}
\end{equation}
where
\begin{equation}
   d\sigma^{p+p\to f+X}
      =\sum_{ij,k...}f_{i/p}(x_1,Q^2)
     \otimes f_{j/p}(x_2,Q^2) \otimes 
     \hat\sigma_{ij\to f+k...} \, .
      \label{eq2.2}
\end{equation}
Here,  $\otimes$ denotes convolutions.
The long-distance information is contained in the parton distributions
$f_{i/p}(x,Q^2)$ of partons $i$ contained in the incoming proton, and in the fragmentation
function $D_{f\to h}(z,\mu^2_F)$ for a parton $f$ to fragment into a hadron $h$.  
The hard partonic cross section
$\sigma_{ij\to f+k...}$ depends in general on the partonic center of mass energy, the
momentum transfer $Q$, the renormalization scale $\mu$, and possibly the masses
of the partons involved in the process. This partonic cross section 
 is calculable perturbatively as a power series in the strong coupling constant 
 $\alpha_s(\mu)$ up to a remainder term,
 which is suppressed by a power of $Q$. The controlled computability of hadronic cross 
sections rests on the scale dependence of the functions in (\ref{eq2.1}) and (\ref{eq2.2}). 
This scale dependence is governed by QCD evolution equations. Physically, it emerges from the
fact that the amount of parton branching, which needs to be included in the 
incoming parton distributions or outgoing fragmentation functions, depends on 
the scale, at which the hard process is interfaced with these functions. 

Equation (\ref{eq2.2}) denotes the single inclusive cross section of hard partons $f$, which
underlies the calculation of jet cross sections. For a jet definition which can be
compared to data, one needs to understand how the partons $f$ fragment, and one requires 
an operational procedure which relates hadronic fragments to jets. In the present chapter,
we discuss jet measurements with an emphasis on those features which are sensitive to
propagation through dense QCD matter. In chapter ~\ref{sec3}, we then turn to 
single inclusive hadron spectra.

 %%%%%%%%%%%%%%%%%%%%%%%%%%%%%%%%%%%%%%%%%%%%%
 \subsection{Parametric estimates relevant for embedding hard processes in the medium}
 \label{sec2.1}
 
 \noindent
 {\bf Factorization.}
 Does factorization apply if one embeds a hard process such as (\ref{eq2.1}) in
 hot and dense QCD matter? Factorization theorems, which 
 include the medium-modifications relevant for heavy-ion collisions, are not known.
 A simple parametric estimate may illustrate the reason: Let us assume that some of the
 partons flowing into or out of the hard process participate in a second interaction with some 
 momentum transfer $Q'$. The cross section of this secondary scattering is of 
 order $\sim \alpha_s/Q'^2$. Hence, medium-modifications are suppressed 
 by powers of $1/Q'$. However, factorization in the sense of (\ref{eq2.1}), (\ref{eq2.2}) 
 is established typically only
 up to terms which are power-suppressed in $1/Q$. Moreover, if the medium dependence 
 results from relatively soft, small-$Q'$ secondary interactions, then 
 the leading term in an expansion of inverse powers of $\alpha_s/Q'^2$ may be
 unreliable. 
  
 For some special classes of measurements in electron-nucleus and
 proton-nucleus collisions, the Liu-Qiu-Sterman (LQS) formalism guarantees the 
 extension of factorization theorems to specific medium effects~\cite{Luo:1994np,Luo:1993ui}. 
 The key observation in
 this context is that there is a parametrically dominant class of geometrically enhanced
 medium-effects, which is proportional to the in-medium path length and hence to $A^{1/3}$.
 These medium effects are of order $\sim \alpha_s\, A^{1/3} /Q'^2$, and they are perturbatively
 calculable. In these cases, hadronic cross sections are known to be given up to an accuracy 
 $O( \alpha_s\, A^{1/3} /Q^2)$ by the convolution of process-independent (twist-four) parton 
 correlation functions in the nucleus  and a hard matrix element. However, the central ideas of the 
 LQS formalism have not been shown to carry over to hadronic cross sections in 
 nucleus-nucleus collisions. 

\noindent
{\bf Localization of the hard process.}
In the phenomenological praxis of heavy ion collisions, one largely bypasses the issue of 
factorization. One aims at constraining with experimental data the absolute 
yield of highly energetic partons produced within some kinematical range in a nucleus-nucleus 
collision, and one then proceeds to calculating the medium-modifications of the propagation and
fragmentation of these partons. In this way, one circumvents  first principles calculations of 
{\it absolute hadronic yields} in the presence of hot and dense QCD matter, for which factorization
theorems would be needed, and one turns to the analysis of  {\it relative hadronic yields};
the absolute yield of the produced hard partons is constrained by data rather than 
calculated. 

\noindent
{\bf Tagged measurements} are the simplest example for this procedure.  In a tagged
measurement, one measures for 
instance a sample of photons or Z-bosons at sufficiently high transverse momentum. 
These  gauge bosons are produced in hard processes but they carry no color charge and
they thus leave the medium unattenuated. The large transverse momentum $p_T$ implies 
that the hard interaction is localized within a small  length scale $\Delta x \sim 1/p_T$
 and time scale $\Delta t \sim 1/p_T$. If this scale is much smaller than the typical length scales
in the medium [which can be expected to be set by the temperature $T \sim O(200\, {\rm MeV})$ or the saturation scale $Q_s\sim O(2\, {\rm GeV})$],
\begin{equation}
\Delta x\, \sim 1/p_T \ll 1/T\, ,\, 1/Q_s\, ,
\label{eq2.3}
\end{equation}
then the hard partonic production process can be expected to be unaffected by the
medium. That means in particular, that the recoil transverse momentum associated with
the triggered Z-boson or photon leaves the hard partonic interaction in the same partonic
configuration as in a proton-proton collision. Conceptually, this turns the recoil partons of
photons and Z-bosons into probes, whose initial energy is constrained kinematically on an 
event-by-event basis, but which can interact with the medium in their subsequent propagation. 
More generally, even if  there is no factorization theorem which determines
the rate of high energy partons produced in hard collisions,
there are multiple and mutually consistent ways of selecting in nucleus-nucleus
collisions data samples, for which the rate of high energy partons produced in the 
initial hard processes is constrained. In general, these high energy partons, after emerging
from the hard interaction, propagate through the medium. To the extent to which the numerical 
size of the resulting medium-effects is much larger than the uncertainties in constraining the
initial kinematical condition of the propagating parton, this enables a characterization
of medium-modifications without relying on theoretical control over the absolute spectra
of hard processes. 

\noindent
{\bf Lifetime of the virtual parton.}
A parton emerging from a hard interaction with high transverse momentum $p_T$ carries an
initial virtuality $Q$, which can be assumed to be logarithmically distributed between $p_T$
and a lower hadronic scale. For high $p_T$, this virtuality is perturbatively large. According to
Heisenberg's uncertainty relation, the parton will degrade its virtuality (i.e. will start evolving closer
to on-shell conditions)  on a time scale $1/Q$ in its rest frame. 
This degradation of virtuality is achieved by
multiple parton branchings and it is at the basis of the final state parton shower. Given that the time-
scale $1/Q$ is small, the question arises to what extent the final state parton shower can be
altered by a medium, and to what extent a branching process will be completed prior to time 
scales of order
$1/T$ or $1/Q_s$, on which medium-effects are expected to become relevant, see eq. (\ref{eq2.3}). 
In this context, it is crucial that the lifetime of the virtual parton is Lorentz-dilated in the rest
frame of the medium by a gamma factor $\sim E_{\rm parton}/m$, where the mass of the parton can be
identified with its virtuality, so
\begin{equation}
	\tau_{\rm virtual\,  life} \simeq \frac{E_{\rm parton}}{Q^2}\, .
	\label{eq2.4}
\end{equation} 
In the extreme case of a parton with $E_{\rm parton}= 100\, {\rm GeV}$  and maximal virtuality
$Q^2 \simeq \left(100\, {\rm GeV}\right)^2$, this time scale is 
$\tau_{\rm virtual\,  life} = (1/ 500)\, {\rm fm/c}$, which is much smaller  than the typical 
wavelengths within the medium components. So, this parton can be expected to branch
prior to interacting with the medium. On the other hand, if a parton with this energy has a 
rather small virtuality of $Q^2 \simeq \left(1\, {\rm GeV}\right)^2$, then
$\tau_{\rm virtual\,  life} = 20\, {\rm fm/c}$. Such a parton will not branch within the path length
of the medium, except if it interacts inelastically with the medium. These numerical 
estimates of hadronization times are indicative of the expected order of magnitude, but 
may vary depending on model assumptions~\cite{Wang:2003aw}.

In general, the description of fragmentation and hadronization in elementary collisions
(i.e. in the vacuum) requires momentum space information only. In contrast, the
description of medium effects must also be based on a picture of how the hard process is
embedded in and propagates through the spatio-temporal region over which the medium 
extends. Time scales of the type (\ref{eq2.4}) determine the spatio-temporal embedding of hard processes and thus decide whether and at what stage within their fragmentation process,
remnants of high energy partons emerging from a hard process will interact with the medium.
We note that at least some aspects of this picture are testable. In particular, 
the non-perturbative stage of parton fragmentation, a.k.a. hadronization, occurs as
soon as the parton shower has evolved to sufficiently low virtuality, say 
$Q^2 \leq \left(1\, {\rm GeV}\right)^2$. According to eq. (\ref{eq2.4}), if at that scale in the
evolution a leading parton still carries an energy of $E_{\rm parton} = 10$ GeV, then
its lifetime is $\tau_{\rm virtual\,  life} \simeq 2$ fm/c. For hadronization to occur 
at $Q^2 \leq \left(700\, {\rm MeV}\right)^2$, we would extract a lifetime of 
$\tau_{\rm virtual\,  life} \simeq 4$ fm/c. Qualitatively, these estimates indicate that if the leading
parton in a parton shower carries more than $O(10)$ GeV energy at the end of the perturbative
evolution, then it can be expected to live long enough to hadronize outside the medium in the 
vacuum. Assuming that leading partons fragment into leading hadrons, this indicates 
that above $p_\perp \geq O(10)$ GeV medium-modifications of single inclusive hadron 
spectra should show the same relative hadrochemical composition 
as in the vacuum, since their hadronization is unaffected by the medium. This is consistent
with observations of the particle-species independence of high-$p_T$ hadron suppression,
made at RHIC. 

\noindent
{\bf Formation time.} The process of parton fragmentation involves numerically important 
quantum interference effects, see section~\ref{sec4}. 
As a result, the picture of a parton shower as a probabilistic 
iteration of parton branching processes has limitations. Interestingly, the most important
interference effects can be included in a probabilistic language. In particular, there is
a one-to-one correspondence which maps the destructive interference between 
subsequent gluon emission onto an angular ordering prescription between 
subsequent probabilistically iterated branchings~\cite{Ellis:1991qj,Dokshitzer:1991wu}. 
This is implemented in technically different ways in the QCD parton showers of 
state of the art Monte Carlo event generators~\cite{Sjostrand:2006za,Corcella:2000bw,Gleisberg:2003xi}.

For targets of finite spatial extension, this is not the only important quantum interference
effect. If a parton branches a gluon before entering a secondary interaction, the question arises 
whether the two decay products should be considered as independent projectiles, or whether
they scatter coherently. A simple quantum mechanical argument helpful in deciding this question
is based on estimating the phase difference between the decay products. Here, the 
relevant phase factor in the wave function of each parton is 
$\sim \exp\left[i E_\perp\, \Delta z \right]$, where $E_\perp$ is the transverse energy and 
$\Delta z$ is the distance in the direction in which the projectile propagates. 
For instance, if a quark fragments $q \to q\, g$ a gluon with energy $\omega$ and transverse 
momentum $k_T$, then the inverse transverse energy defines the formation time
\begin{equation}
  \tau_{\rm form}  = \frac{2\omega}{k_T^2}\, .
   \label{eq2.5}
\end{equation}
For $\Delta z = \tau_{\rm form}$, the relative phase $E_\perp\, \Delta z$ between the
two daughter parton wave functions is unity, and that indicates decoherence of the wave 
function of the emitted gluon from its parent. On time scales small compared to $\tau_{\rm form}$,
however, the two components of the projectile wave function can be expected to act coherently,
that means, scattering proceeds as if the gluon is not yet formed, see section~\ref{sec4.4.3}
for more details.

We note that if a virtual parton splits in the vacuum into two approximately massless
daughters with momentum fractions $z$ and $(1-z)$, then the relative transverse
momentum between the outgoing partons satisfies $k_T^2 \simeq z\, (1-z)\, Q^2$.
Taking the soft daughter parton to be the gluon with energy $\omega = z\, E_{\rm parton}$
and setting $(1-z) \simeq 1$, one finds $2\, \omega / k_T^2 \simeq 2\, E_{\rm parton}/Q^2$.
This shows that the estimates (\ref{eq2.4}) and (\ref{eq2.5}) are closely related.  
%
%%%%%%%%%%%%%%%%%%%%%%%%%%%%%%%%%%%%%%%%%%
%
%
\boldmath
\subsection{Jet definitions}
\label{sec2.2}
\unboldmath
A jet is the collimated set of hadronic decay products of a parent parton.
But the concept of a parton is ambiguous, for instance because it is scale dependent.
As a consequence, there are different definitions of what a jet is, and these correspond to
different algorithms for searching jets in hadronic collisions and - in general - they yield
different results. It is hence important that the theoretical calculation of a jet
matches the measurement procedure. According to the {\it SNOWMASS accords}~\cite{snowmass}, which were agreed on by experimentalists 
and theorists in 1990, a 'good' jet definition should be simple to implement in experimental
analysis and theoretical calculations, it should be defined and yield finite cross sections
at any order of perturbation theory, and the obtained cross sections should be relatively
insensitive to hadronization. Because of the requirement on the validity of perturbation 
theory, the algorithm with which jets are characterized in hadronic collisions must be 
infrared and collinear safe, i.e., the notion of a jet should not depend upon adding or
subtracting a soft particle or upon collinearly splitting a hard particle. Also, it should
be as insensitive as possible to the underlying event. 

There has been significant progress recently on improving jet definitions
(see e.g. Ref.~\cite{Soyez:2008su} and references therein)
in line with the SNOWMASS accords and there are arguments that this novel generation
of jet definitions should be suited for analyzing jets in the high-multiplicity environment
created in heavy ion collisions. So far, essentially all high-$p_T$ data analyses at RHIC have 
been carried out either 
on the level of single inclusive hadron spectra, or by constructing jet-like particle correlations
with the help of high-$p_T$ trigger particles. Only recently, a first attempt was made to
use calorimetric jet definitions for the
analysis of RHIC data~\cite{Salur:2008hs}. This analysis identified calorimetric 
towers with a jet energy $E_{\rm jet} \leq 40\, {\rm GeV}$, which is large for RHIC kinematics, 
but which is still small compared to the energy scales at which one characterizes
jets normally. These kinematical limitations at RHIC are certainly one of the reasons for
why there is still little experience in applying calorimetric jet definitions  to the high-
multiplicity environment of heavy ion collisions. We expect that this situation will change
radically with the much wider kinematic range accessible at the LHC.

%%%%%%%%%%%%%%%%%%%%%%%%%%%%%%%%%%%%%%%%%%
\subsubsection{Jet algorithms}
\label{sec2.2.1}
Classical jet algorithms can be grouped roughly into two classes:\\
{\bf Cone algorithms} aim at defining jets as dominant directions of energy flow within
a circle of radius 
\begin{equation}
 R = \sqrt{ (\Delta y)^2 + (\Delta\phi)^2}
  \label{eq2.6}
\end{equation}  
in the $(y,\phi)$-plane of rapidity $y$ and azimuth $\phi$.  [Longitudinally invariant
cone algorithms are clearly formulated in terms of rapidity rather than pseudo-rapidity $\eta$,
but there are cone algorithms formulated in $y$.] The center of this cone is 
defined by the sum of the momenta lying in the cone. Most cone algorithms are 
seeded, that means that the algorithmic reconstruction of jets starts from a set of seeds
which are e.g. all calorimetric towers in the $(y,\phi)$-plane showing more than a
certain energy. The idea is then to vary the content of the cone and thus its direction, 
till it coincides with a local maximum of energy flow in the $(y,\phi)$-plane. 
A complication of cone algorithms is that different cones may overlap. One thus
requires a prescription which either prevents the algorithm from finding overlapping cones,
or which defines how to distribute the content in the overlap of two jet cones. Both avenues
have been explored in the praxis. In particular, by iteratively removing those calorimetric 
towers from the event, which have been attributed to a jet (iterative cone algorithms with 
progressive removal), one ensures that overlap does not occur. Alternatively,
there are iterative procedures which amount to splitting and merging overlapping cones
(cone algorithms with split-merge). A second complication is that for the simplest jet cone 
algorithms, the addition of an infinitely soft particle can in principle change the clustering 
of jets in an event. Most currently used cone algorithms have procedures implemented
to control or remedy this phenomenon, but the performance of these procedures in 
the high-multiplicity environment of a heavy ion collision deserves further studies. 
There is one recently proposed cone algorithm, \textsc{SISCONE}~\cite{Salam:2007xv}
(Seedless Infrared-Safe Cone jet algorithm) which meets the SNOWMASS accords, 
that means, it is infrared and collinear safe. 

From jet production at LEP, LEP2 and Tevatron, one know that for jets of energy 
100 GeV, approximately 70 \% of the jet energy is contained in a cone of radius 
$R=0.3$, and approximately 90 \% of the jet energy is contained in a cone of 
radius $R= 0.5$.  These jet energy fractions narrow slightly with increasing jet energy, 
and there is a parametrization based on data from the D0-collaboration~\cite{Abbott:1997fc}. 
Clearly, to capture most of the jet energy, a cone radius $R>0.3$ is needed. 
In hadronic collisions, one typically uses cone radii $0.7 < R < 1.0$. 

\noindent
{\bf Jet reconstructions based on successive recombinations.}
These reconstruction algorithms are based on defining a distance $d_{ij}$ between any pair of
objects in an event as well as a so-called beam distance $d_{iB}$ for each object. For each event,
one identifies the smallest distance. If this distance is smaller than the beam distance, then
one combines the two objects into one. If it is larger, then the object is called a jet and removed
from the event. This is repeated till no object is left. There are essentially three successive
recombination algorithms, which are infrared and collinear safe. They are defined by
the distance measures
\begin{eqnarray}
	d_{ij} &=& {\rm min}\left(k_{T,i}^{2p}, k_{T,j}^{2p} \right)\, 
		\left( \Delta y_{ij}^2 +  \Delta \phi_{ij}^2 \right)\, ,
		\label{eq2.7}
		\\
	d_{iB} &=& R^2\,  k_{T,i}^{2p}\, .
		\label{eq2.8}
\end{eqnarray}
Depending on the integer $p$, these distances define the Cambridge/Aachen 
algorithm~\cite{Dokshitzer:1997in,Wobisch:1998wt} for $p=0$, the $k_T$ 
algorithm~\cite{Catani:1993hr}  
for $p=1$ and the anti-$k_T$  algorithm~\cite{Cacciari:2008gp} for  $p=-1$. 

A variant of the $k_T$ algorithm specially suited for $e^+\, e^-$ collisions 
is the Durham clustering algorithm~\cite{Catani:1991hj}, where one defines
for each pair of final state particles the distance 
\begin{equation}
d_{ij} = 2 {\rm min}(E_i^2,E_j^2) (1-\cos \theta_{ij}) / E_{\rm cm}^2\, .
\label{eq2.16}
\end{equation}
The pair of particles with smallest $d_{ij}$ is then replaced by a
pseudo-particle, whose energy and momentum are the sums of its daughters. The 
procedure is repeated until all $d_{ij}$ exceed a given threshold $d_{\rm cut}$.

In the above, we have focussed mainly on jet algorithms of the second generation, which
meet the SNOWMASS accords. A short overview of other currently used jet algorithms 
is given in Ref.~\cite{Soyez:2008su}.

%%%%%%%%%%%%%%%%%%%%%%%%%%%%%%%%%%%%%%%%%
\subsubsection{Background and background fluctuations for jet reconstruction}
\label{sec2.2.2}
In a central Pb-Pb collision at the LHC, there is - unrelated to jet production -
typically a total transverse energy per unit rapidity of $dE_T/d\eta$ of 1 TeV or more. 
This estimate is obtained by multiplying the expected average 
$\langle \sqrt{p_T^2}\rangle \sim 700$ MeV per hadron with a minimal multiplicity 
$dN/d\eta = 1500$ of the sum of charged and neutral particles. The area 
$A_{\rm jet} = \pi\, R^2$ of a jet cone covers a significant fraction of the entire 
area $A_{\rm total} = \Delta \eta \times \Delta \phi = 2\, \pi$ within one unit of 
pseudo-rapidity. As a consequence, the high-multiplicity environment leads 
- unrelated to jet production - to an energy of 
\begin{equation}
	E_{bg}(R) > \frac{A_{\rm jet}}{A_{\rm total}} 1\, {\rm TeV}
	= \Bigg\lbrace 
	\begin{array}{c} 45\,  {\rm GeV}\quad \hbox{for}\, R = 0.3 \\ 
	125\,  {\rm GeV}\quad \hbox{for}\, R = 0.5 \\
	245\,  {\rm GeV}\quad \hbox{for}\, R = 0.7
	\end{array}
	\label{eq2.9}
\end{equation}
These numbers indicate that in a heavy ion collision, the transverse energy of the
underlying event in a cone of typical radius $0.3 < R < 0.7$ is comparable in magnitude
to the energy added by the jet. Reducing the 
cone size or applying a transverse momentum cut does both reduce the background
strongly whereas it affects the jet signal to a much lesser extent.

The reconstruction of the energy of a jet can be at best as accurate as the estimate of the 
background energy, which is contained in the same jet cone and which must be subtracted. 
For event-by-event jet reconstruction, this accuracy is limited by the fluctuations in the 
background. We distinguish two different contributions:
\begin{enumerate}
	\item {\it Fluctuations caused by event-by-event variations in impact parameter}\\
	A centrality class is a selection of events with a certain spread in multiplicity and in
	 impact parameter.  This spread translates into a fluctuation of the total background
	 energy inside a cone
	 \begin{equation}
	 		\Delta E_{\rm bg}^{\rm EbyE} \propto R^2\, .
		\label{eq2.10}
	 \end{equation}
	 Since the variation of impact parameter affects multiplicity inside and outside the
	 cone in a correlated way, information from outside the cone (e.g. from other
	 rapidity windows) can be used to estimate the effects of this fluctuation on an
	 event-by-event basis.
	 \item {\it Out-of-cone multiplicity fluctuations in events at fixed impact parameter}\\
	 Under the assumption that the particles produced in the collision are uncorrelated,
	 these background fluctuations are Poissonian, and their r.m.s. is
	 (see e.g. Ref.~\cite{Carminati:2004fp})
	 \begin{equation}
	 		\Delta E_{\rm bg}^{\rm Poisson} 
				= \sqrt{N}\, \sqrt{ \langle p_T\rangle^2 + \sigma_{p_T}^2}
			\propto R\, .
	     \label{eq2.11}
	 \end{equation}
	Here, $N$ is the number of uncorrelated particle in the cone, and $\sigma_{p_T}$
	is the r.m.s. of the transverse momentum spectrum. However, the intermediate and 
	high-$p_T$ particles, which dominate $\Delta E_{\rm bg}^{\rm Poisson} $, show 
	jet-like correlations. This leads to fluctuations which are stronger than those
	obtained in the Poissonian limit
		 \begin{equation}
	 		\Delta E_{\rm bg}^{\rm real} > \Delta E_{\rm bg}^{\rm Poisson} 
			\label{eq2.12}
		 \end{equation}
	Any estimate about the extent to which the Poisson assumption underestimate
	multiplicity fluctuations requires a dynamical understanding of jet-like correlations. 
\end{enumerate}

\boldmath
\subsection{Characterizations of the intra-jet structure}
\label{sec2.3}
\unboldmath
Since jets are multi-particle final states, a large number of independent measurements
has been used for their characterization. In general, these measurements characterize
the energy flow, the particle distribution, particle correlations and particle identity within
a jet. In the following, we discuss the most common measures. 
%

%%%%%%%%%%%%%%%%%%%%%%%%%%%%%%%%%%%%%%%%%%%%
\subsubsection{Jet event shapes}
\label{sec2.3.1}
Some of the best studied measurements of jet energy flow correspond to perturbatively 
calculable, infrared-safe quantities. In the context of heavy-ion physics, their study has
started only recently, but it is of potential interest for several reasons. First, perturbative 
calculability indicates that
the measurement is mainly determined by the partonic large-$Q^2$ part of the parton shower. 
This corresponds to very early times of order $\sim E/Q^2$ into the parton shower evolution 
[see equation (\ref{eq2.4})], and thus any medium-modification of such measurements can 
be expected to be sensitive to the early and dense stage of the collision. Second, 
infrared-safe quantities may be easier to characterize within the high-multiplicity 
environment of a heavy-ion collision, since infrared safety implies a relatively weak
dependence (ideally, an insensitivity) to the underlying event. 
%
%%%%%%%%%%%%%%%%%%%%%%%%%%%%%%%%%%%%%%%%%%%%%%%%%%%%%%%%%%%%%%%%%%%%
\begin{figure}[h]\epsfxsize=7.9cm
\centerline{\epsfbox{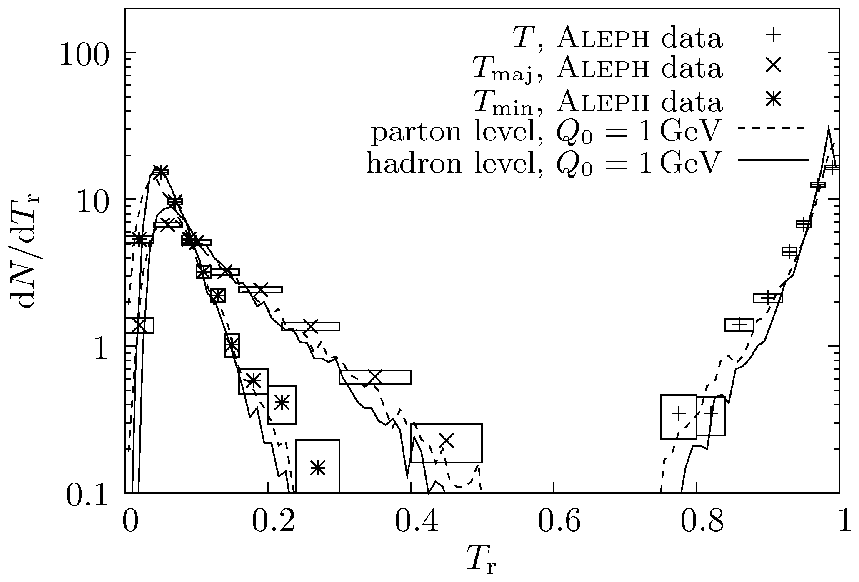}
\epsfbox{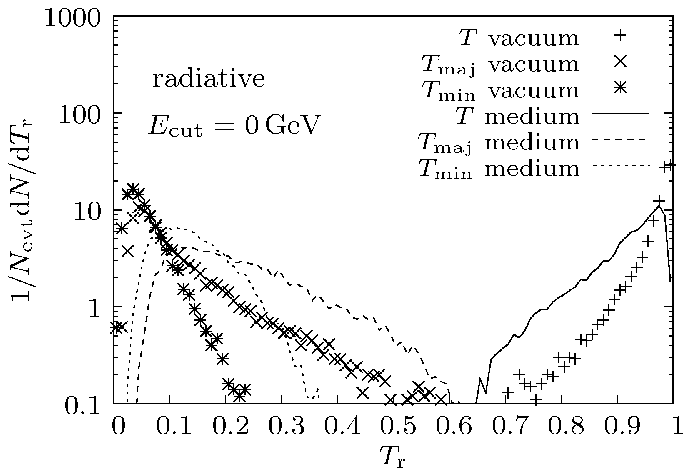}}
% \centerline{\epsfbox{mediummodif.eps}}
%\vspace{0.5cm}
\caption{The thrust, thrust major and thrust minor ($T_\text{r}=(T,T_\text{maj},T_\text{min})$) 
distributions for jets in $\sqrt{s}=\unit[200]{GeV}$ $e^+e^-\to q\, \bar{q}\to X$
collisions. LHS: Data of the ALEPH Collaboration~\cite{Heister:2003aj} 
are compared to a Monte Carlo simulation with and without subsequent hadronization.
RHS: In a model, in which the medium induces additional gluon radiation, all thrust 
distributions broaden significantly compared to the vacuum baseline. Figures taken
from Ref.~\protect\cite{Zapp:2008gi}.
}\label{figthrust}
\end{figure}
%%%%%%%%%%%%%%%%%%%%%%%%%%%%%%%%%%%%%%%%%%%%%%%%%%%%%%%%%%%%%%%%%%%
%
Thrust $T$, thrust major $T_{\rm maj}$ and thrust minor $T_{\rm min}$ are amongst
the best studied perturbatively calculable, infrared-safe jet event shapes~\cite{Ellis:1991qj}. 
For these quantities, one sums over the three-momenta $\vec{p}_i$ of all final state
particles. According to the definition of thrust,
\begin{equation}
	T \equiv {\rm max}_{\vec{n}_T} 
	\frac{\sum_i \vert \vec{p}_i\cdot \vec{n}_T\vert}{\sum_i \vert
\vec{p}_i\vert}\, ,
	\label{eq2.13}
\end{equation}
a 2-jet event is pencil-like if $T=1$, that is if all particles are aligned
parallel or antiparallel to a thrust axis $\vec{n}_T$. The event is spherical if $T=1/2$.  
Thrust major and thrust minor characterize the jet energy flow in the plane orthogonal 
to the thrust axis $\vec{n}_T$. Thrust major is defined  as the projection
of all particle momenta on the direction $\vec{n}$, which is orthogonal to
$\vec{n}_T$ and along which the momentum flow is maximal
\begin{equation}
	T_{\rm maj} \equiv {\rm max}_{\vec{n}_T\cdot \vec{n}=0} 
	\frac{\sum_i \vert \vec{p}_i\cdot \vec{n}\vert}{\sum_i \vert
\vec{p}_i\vert}\, .
	\label{eq2.14}
\end{equation}
Thrust minor sums up the components $\vec{p}_{ix}$ of the final particle
momenta $\vec{p}_{i}$, which are  orthogonal to the plane defined by
$\vec{n}$
and $\vec{n}_T$,
\begin{equation}
	T_{\rm min} \equiv 
	\frac{\sum_i \vert \vec{p}_{ix}\vert}{\sum_i \vert \vec{p}_i\vert}\, .
	\label{eq2.15}
\end{equation}
The left hand side of Fig.~\ref{figthrust} shows data from  the ALEPH collaboration, 
compared to a Monte Carlo simulation. We note that according to perturbative
QCD, the destructive interference between successive gluon emissions translates
into a strong angular ordering of the parton shower. As a consequence, the first
branching process in the parton shower typically carries significantly more 
transverse momentum (measured with respect to the thrust axis $\vec{n}_T$)
than subsequent branchings. This implies that the first 
branching largely determines thrust major. Also, the orientation of the second
branching with respect to $\vec{n}_T$ and $\vec{n}$ significantly influences
the degree to which thrust minor is more narrow than thrust major. On the
level of these qualitative considerations, one sees already that thrust, thrust
major and thrust minor provide a detailed test of the dynamical description of
multiple parton branching processes. Moreover, since branchings at low virtualities 
$Q^2$ do not result in significant transverse momenta between the daughter
partons, these quantities are relatively insensitive to the "late" low-$Q^2$ stage 
of the parton shower and to hadronization. The right hand side of Fig.~\ref{figthrust}
illustrates that despite their perturbative calculability for elementary interactions,
these event shapes show in model calculations 
a strong sensitivity to a class of potential medium-modifications.
For instance, any interaction between the medium and the parton shower, which 
induces additional gluon radiation at early stages of the parton shower does leave
distinct traces in thrust, thrust major and thrust minor. 

The interaction of a jet with the dense QCD matter produced in heavy 
ion collisions can be expected to lead to a broadening of the jet energy flow,
which can be characterized by jet shape observables. Characteristics 
of the broadening can be expected to be sensitive not only to properties of the dense
QCD matter, but they will also give access to the microscopic dynamics of the 
interaction between probe and medium. On general grounds, for instance,
elastic interactions between probe and medium (which are sometimes
referred to as collisional energy loss) are dominated by small-angle
scattering processes, and the degrees of freedom in the medium which accept
the scattering recoil, will have low energy in comparison to the projectile partons. 
As a consequence elastic interactions are likely to lead to a broadening of jet shape 
observables, which is dominated by low-$p_T$ particles. In contrast, any 
medium-induced additional parton splitting tends to increase the broadening of
jet event shapes by modifying the parton shower in the range of intermediate and
high transverse momenta. There are by now first quantitative model 
studies which support this qualitative idea~\cite{Zapp:2008gi}.

There is a wide class of event shape observables.
It includes quantities such as oblateness, sphericity, planarity,
aplanarity and total jet broadening. In principle, many of these observables are 
independent. In the practice of comparing Monte Carlo simulations of multi-particle
final states with data,  it turns out that simulations typically account 
satisfactorily for other event shape observables, if they account for 
thrust, thrust major and thrust minor. For this reason,  we refer to the literature for 
discussion and data of other event jet observables.  

%%%%%%%%%%%%%%%%%%%%%%%%%%%%%%%%%%%%%%%%%%%%%%%
\subsubsection{Jet substructures}
\label{sec2.3.2}
As discussed in section ~\ref{sec2.2}, the definition of jets depends on the distance
scale with respect to which jets are defined. This distance can be a cone size $R$,
see eq. (\ref{eq2.6}), or a distance of the type (\ref{eq2.7}), (\ref{eq2.8}). 

To be specific, consider the Durham cluster algorithm (\ref{eq2.16}). 
The number of clusters separated by a distance larger than $d_{\rm cut}$ is
defined to be the number $n$ of jets. 
%
%%%%%%%%%%%%%%%%%%%%%%%%%%%%%%%%%%%%%%%%%%%%%
% \begin{figure*}[t]
% \centering
% \input{2-njetvac-lqcd300.pstex_t}
% \caption{The jet rates as a function of jet resolution scale $y_{\rm cut}$ in 
% $\sqrt{s}=\unit[200]{GeV}$ $e^+e^-\to q\, \bar{q}\to X$ collisions.
% Left hand side: Simulation of \textsc{Jewel} with and without hadronisation for evolution 
% down to $Q_0=\unit[1]{GeV}$. Right hand side: Data of the \textsc{Aleph}
% collaboration~\cite{Heister:2003aj} compared to simulations of Jewel with hadronisation.
% Figures taken from Ref.~\protect\cite{Zapp:2008gi}. }
% \label{fig2}
% \end{figure*}
%%%%%%%%%%%%%%%%%%%%%%%%%%%%%%%%%%%%%%%%%%%
%
The so-called $n$-jet fraction measures then the number of jets in an event as a function
of the resolution $y_{\rm cut}$. As one decreases the resolution 
scale $y_{\rm cut}$, a parton fragmentation pattern
which has been counted as a single jet for some resolution $y_{\rm cut}$, may be
resolved into more than one jet. This dependence of 
the number of jets in an event on the resolution $y_{\rm cut}$ is particularly sensitive
to the discrete and stochastic nature of the partonic fragmentation process. 

If the interaction of a jet with the QCD matter results in additional, sufficiently energetic,
medium-induced gluon radiation, then these additional partons can be expected to
be seeds of additional jets at sufficiently fine resolution $y_{\rm cut}$. As a consequence,
one expects to count more jets at small $y_{\rm cut}$. On the other hand, elastic 
interactions between jet and medium are unlikely to have a significant effect on
the $n$-jet fraction, since they are dominated by small-angle scattering, which has
little effect on the distance (\ref{eq2.16}). First Monte Carlo studies support these
qualitative conclusions~\cite{Zapp:2008gi}.

%%%%%%%%%%%%%%%%%%%%%%%%%%%%%%%%%%%%%%%%%%%%%%
\subsubsection{Jet multiplicity distributions}
\label{sec2.3.3}
In this subsection, we discuss mainly the understanding of inclusive single-hadron
intra-jet multiplicity distributions $dN/d\xi$, $\xi = \ln \left[ 1/x\right]$ 
as a function of the logarithm of the hadron momentum fraction $x= p/E_{\rm jet}$ 
along the jet axis.  These distributions can be measured for all charged hadrons,
or for identified hadron species. In principle, they can also be measured 
separately for quark and for gluon-initiated jets. In what follows,  we shall denote by 
$dN/d\xi$ not only hadronic, but also partonic multiplicity distributions. Since the
multiplicity in a parton shower increases during the evolution, the latter are functions 
of the evolution scale 
$Y \equiv \ln \left[Q/Q_0 \right]$, $Q_0 \leq O(\Lambda_{\rm QCD})$.

%
%%%%%%%%%%%%%%%%%%%%%%%%%%%%%%%%%%%%%%%%%%%%%%%%%%%%%%%%%%%%%%%%%%%%
\begin{figure}[h]\epsfxsize=10.7cm
%\centerline{\epsfbox{ktmorkap.eps}}
%\centerline{\epsfbox{ktmorom.eps}}
\centerline{\epsfbox{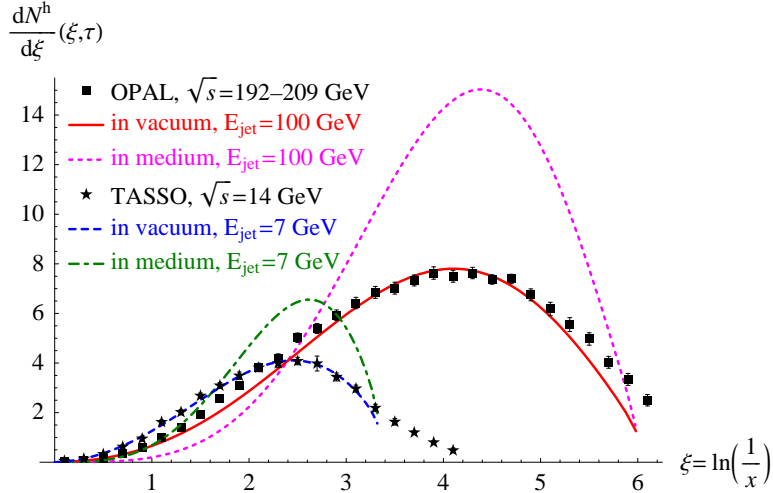}}
%\vspace{0.5cm}
\caption{The single inclusive hadron distribution as a function of 
$\xi = \ln \left[E_{\rm jet}/p\right]$. Data are taken from the $e^+e^-$ collision
experiments TASSO and OPAL,
$E_{\rm jet} = \sqrt{s}/2$. Lines through data denote the MLLA limiting spectrum 
described in the text. Dashed and dash-dotted curves labeled 
"in medium" are calculated for a model in which medium effects enhance gluon emission.
Figure taken from~\cite{Borghini:2005em}.
}\label{fighumpback}
\end{figure}
%%%%%%%%%%%%%%%%%%%%%%%%%%%%%%%%%%%%%%%%%%%%%%%%%%%%%%%%%%%%%%%%%%%
In contrast to jet shapes, which are relatively insensitive to hadronization,
one may expect that the hadronic multiplicity within a jet shows a stronger 
dependence on the dynamics at hadronization, since any hadronic decay 
process affects the hadronic yield in the final state. Remarkably, however,
perturbative QCD has been used with significant phenomenological success
in comparisons of the shape and hadrochemistry of inclusive single-hadron
jet multiplicity distributions~\cite{Dokshitzer:1988bq,Fong:1990nt,Azimov:1984np}. 
Qualitatively, this may be understood as a consequence
of two facts: First, hadronization becomes relevant at low virtuality. This low virtuality
limits the phase space of further branching processes and hence it limit
the degree to which the hadronic yield can be augmented compared to the 
perturbatively calculated partonic multiplicity above but close to hadronization scale. 
Second, there is evidence that important 
aspects of hadronization are local, in the sense that partonic multiplicities give rise 
to hadronic multiplicities within the same phase space region. Taken together, 
these observations support a picture, in which hadronic intra-jet multiplicity 
distributions can be related to partonic multiplicity distributions by an overall fit 
factor of order unity, which does not depend on kinematical variables. Perturbative
QCD can then predict the shape and jet energy dependence of $dN/d\xi$. 

For details of the perturbative description of jet multiplicities, we refer to the 
literature~\cite{Dokshitzer:1988bq,Fong:1990nt}. 
Here, we note solely that 
destructive interference between soft gluon emissions within a jet is known since
the early days of QCD to suppress hadron production at small momentum fractions 
$x= p/E_{\rm jet}$. Already in the double logarithmic approximation, the QCD evolution 
equations in $Y$ show the characteristic hump-backed plateau, seen in data
of $dN/d\xi$, see Fig.~\ref{fighumpback}. For a quantitatively reliable description
of sufficiently small
momentum fractions $x$ and sufficiently large jet energies, where
$\ln\left[1/x \right] \sim \ln \left[E/Q_0 \right] \sim {\cal{O}}(1/\sqrt{\alpha_s})$, however,
one must take into account terms of relative order
$\sqrt{\alpha_s}$. This is achieved in 
in the modified leading logarithmic approximation (MLLA), which provides an 
analytically controlled calculation in the region of sufficiently large $\xi$, only. 
Supplementing this partonic MLLA calculation~\cite{Dokshitzer:1988bq,Fong:1990nt} 
with the hypothesis of local parton hadron duality (LPHD)~\cite{Azimov:1984np} 
results in a good agreement with the 
jet multiplicity distributions observed in elementary collisions, 
 see e.g. Fig.~\ref{fighumpback}.  
 
 While MLLA accuracy per se is not sufficient to describe the 
 inclusive single-hadron distribution as a function of the transverse momentum
 with respect to the jet axis, this is possible in a recent approach called 
 NMLLA~\cite{Ramos:2006dx,Arleo:2007wn},
 in which some parametrically higher-order terms are kept. Physically, these terms
 amount to improving energy momentum conservation at each parton branching.
 Exact energy conservation combined with MLLA accuracy can be achieved by 
 numerical techniques~\cite{Sapeta:2008km} and appears to be quantitatively 
 important in most of the $Y$- and $\xi$-range explored experimentally so far.  
  
The presence of dense QCD matter produced in a heavy ion collision is expected
to degrade the energy of the most energetic partons in the shower. 
Since the jet energy is conserved, a reduction in the
energy of the most energetic partons within a jet implies an increase in the total
jet multiplicity.  Figure~\ref{fighumpback}
shows a model study which illustrates this phenomenon for the case of medium-induced
additional gluon splitting. Most generally, one expects that the yield of high energy 
partons (small $\xi$) decreases while the yield of partons of lower energy (large $\xi$)
increases. Assuming local parton hadron duality, this should be reflected directly
in the hadronic distribution. One also expects that the multiplicity distribution broadens
in transverse momentum space due to multiple scattering. 

%%%%%%%%%%%%%%%%%%%%%%%%%%%%%%%%%%%%%%%%
\subsubsection{Jet hadrochemistry}
\label{sec2.3.4}
From a general perspective, highly energetic partons propagating through a dense
plasma can be viewed as probes, which are initially very far away from the 
equilibrium state of the surrounding matter. By interacting with the medium, they 
participate in equilibration processes, which can by characterized by studying
the medium-modification of jets. The measurements discussed so far allow us
to characterize kinetic aspects of equilibration. In contrast, jet hadrochemistry
addresses the question to what extent the jet embedded in the medium participates 
in hadrochemical equilibration processes. The sensitivity of hadrochemical jet 
measurements arises from the fact that the hadrochemical
composition of jets in the vacuum is known to differ characteristically from that of the
bulk hadronic composition in heavy ion collisions~\cite{BraunMunzinger:2003zd}. 
It also differs from the hadrochemical
composition in RHIC heavy ion collisions  at intermediate transverse momentum, which 
seems to follow quark counting rules~\cite{Abelev:2006jr}. 

From the previous subsection, we know that important qualitative and quantitative 
features of jet multiplicity distributions can be accounted for by perturbation theory. 
In particular, supplementing MLLA ~\cite{Dokshitzer:1988bq,Fong:1990nt} with 
LPHD~\cite{Azimov:1984np}
amounts to extending the perturbative 
shower evolution down to $\Lambda_{\rm QCD}$ and then assuming a one-to-one
correspondence between partonic and hadronic degrees of freedom. For jets in
elementary collisions, this hadronization prescription has been applied successfully
to jet hadrochemistry. By evolving the parton shower for different hadron species 
down to scales set by the hadron masses, one can account for the main characteristic
differences in the hump-backed plateaus of pions, kaons and protons. If this
hadronization assumption persists in the presence of a dense QCD matter, then
it leads to specific predictions for the medium-modified hadrochemical composition
of quenched jets~\cite{Sapeta:2007ad}. 
%
%%%%%%%%%%%%%%%%%%%%%%%%%%%%%%%%%%%%%%%%%%%%%%
\begin{figure}[t]
\includegraphics[scale=.375,angle=0]{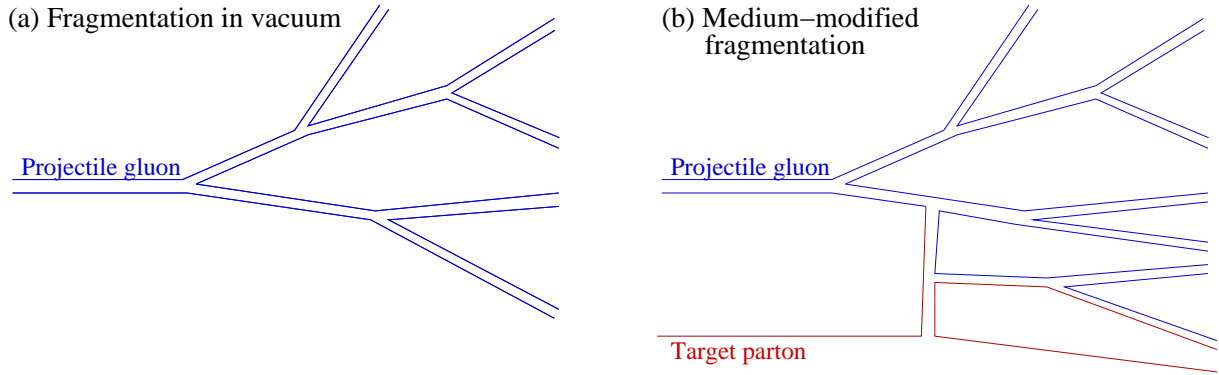}
\caption{Sketch of an entirely gluonic parton shower in the large $N_c$ limit, where gluons
are represented as pairs of $q\bar{q}$ fermion lines, and quarks as single lines.
(a) Fragmentation of the gluon in the vacuum. (b) Interaction of the gluon with a target
quark in the medium via a single gluon exchange. This interaction changes the color flow
and may affect hadronization, see text. 
Figure taken from Ref.~\cite{Sapeta:2007ad}.
}
\label{fighadchem}
\vskip-0.05in
\end{figure}
%%%%%%%%%%%%%%%%%%%%%%%%%%%%%%%%%%%%%%%%%%%%%%%

However, several other dynamical mechanisms are conceivable, which may affect the 
hadrochemistry of jets in the medium considerably. In particular, 
as sketched in Fig.~\ref{fighadchem}, any gluon exchange between the parton shower
and the medium changes the color flow. Since hadronization must implement 
color neutralization, this can change significantly the hadrochemical composition of
jet fragments. In particular, for the case of a hadronization model based on string
fragmentation, Fig.~\ref{fighadchem} illustrates clearly that the initial invariant mass
distribution of strings may be expected to be rather different from that in the vacuum. 
One may also speculate on other mechanisms. For instance, if components of the
medium are kicked by the parton shower to sufficiently high transverse momentum,
then the hadrochemical composition of what is kicked can affect jet hadrochemistry.
We expect that jet hadrochemistry will be characterized in detail within the LHC
heavy ion program.

%
%%%%%%%%%%%%%%%%%%%%%%%%%%%%%%%%%%%%%%%%%%%
\boldmath
\section{Leading hadrons in the absence and in the presence of a medium}
\label{sec3}
\setcounter{equation}{0}
\unboldmath
At sufficiently high transverse momentum, inclusive single-hadron spectra in proton-proton
collisions can be calculated within the perturbative QCD factorized formalism, as discussed
shortly in the context of equations (\ref{eq2.1}) and (\ref{eq2.2}). The qualitative 
discussion of length scales and time scales in section~\ref{sec2.1} applies to 
high-$p_T$ inclusive hadron spectra, as well. In particular, high-$p_T$ hadrons result from 
highly energetic partons, which have evolved in the parton shower down to a hadronic
virtuality $Q_{\rm hadr}$. The lifetime of these partons can be estimated according to
(\ref{eq2.4}) to be of the order $E_{\rm parton}/ Q_{\rm hadr}^2$, which takes values of
the order of a nuclear radius for $E_{\rm parton}\simeq 10$ GeV and typical hadronization
scales. The far-reaching consequence of this parametric estimate is that for sufficiently
high transverse momentum, leading hadrons form outside the medium. Thus, the
medium-modification of high-$p_T$ inclusive hadron spectra is expected to be 
sensitive solely to the medium-modification of parton propagation, and should not depend
on the interaction of hadronized fragments with the medium. 
An important test of this conclusion is to verify that the particle-secies dependence
of medium-modifications is trivial, i.e., that it arises solely  from differences in
the medium-modification of the parent gluons, light quarks and heavy quarks. 

%%%%%%%%%%%%%%%%%%%%%%%%%%%%%%%%%%%%%%%%%%%
\boldmath
\subsection{Trigger biases}
\label{sec3.1}
\unboldmath
Hadronic measurements select classes of partonic fragmentation patterns. 
Depending on the measurement, this selection can show a significant bias.
These trigger biases can be seen clearly, for instance, on the level of
the average momentum fraction $\langle z \rangle$, which is carried by the most
energetic hadron inside the experimentally selected partonic fragmentation patterns.

\noindent
{\bf Trigger bias in the vacuum.} The calculation of single inclusive spectra proceeds
by convoluting the probability $D_{f\to h}(z)\, dz$ of a parton $f$ of transverse momentum
$p_T$ to fragment into a hadron $h$ of momentum fraction $z = p_T^h / p_T$
with the probability $\frac{d\sigma^{p+p\to f+X}}{dp_T} dp_T$ of producing a
parton with this momentum. Let us approximate the partonic cross section by
a power-law $ \frac{d\sigma^{p+p\to f+X}}{dp_T} \propto \frac{1}{\left(p_T\right)^n}$.
It then follows from the hadronic cross section 
$d\sigma^{p+p\to h+X} = \frac{d\sigma^{p+p\to f+X}}{dp_T}\, dp_T\, D_{f\to h}(z)\, dz$
that its dependence on hadronic momentum $p_T^h$ is determined by the
$(n-1)$-th moment of the fragmentation function
\begin{equation}
	\langle z \rangle_{lead.hadr.} \simeq \int z^{n-1}\, D_{f\to h}(z)\, dz\, .
	\label{eq3.1}
\end{equation}
Within the kinematical ranges tested at RHIC and at the LHC, the hard partonic cross
section is significant steeper than the $n=4$ dependence expected of lowest 
order perturbative parton-parton scattering (which should be valid in the limit
$p_T \to \infty$ with $s/p_T^2 = {\rm fixed}$). One of the main reasons for this
steeper fall-off of  $ \frac{d\sigma^{p+p\to f+X}}{dp_T} $ with increasing $p_T$
is the $x$-dependence of the parton distribution functions entering (\ref{eq2.2}).
These decrease with increasing transverse parton momentum $p_T$, since
$x_1\, , x_2 \propto \sqrt{p_T^2/s}$. There are also other effects, for instance,
the running of the coupling constant $\alpha_s(p_T^2)$ in the hard parton-parton interaction 
adds to increasing the power $n$. Clearly, for quantitative statements about the 
$p_T$-dependence of $ \frac{d\sigma^{p+p\to f+X}}{dp_T} $, one should turn
to a reliable perturbative calculation. For the purpose of this argument, however,
it is sufficient to note that one has $n \geq 7$ over a wide kinematical range at RHIC 
and at the LHC.

Let us contrast the measurement of a single-inclusive hadron spectrum with a 
measurement, in which one selects a set of events containing partons $f$ of
known transverse momentum $p_T$. In fact, calorimetric jet measurements aim
at selecting such an ideal event sample by measuring jets with energy $E_{\rm jet} = p_T$.
According to the definition of a fragmentation function, the most energetic hadron in 
these jets carries an average momentum fraction
\begin{equation}
	\langle z \rangle_{jet} \simeq \int z\, D_{f\to h}(z)\, dz\, .
	\label{eq3.2}
\end{equation}
By comparing equations (\ref{eq3.1}) and (\ref{eq3.2}), one finds that 
single inclusive hadron spectra amount to a dramatic trigger bias, since the 
$(n-2)$-nd moment of the fragmentation function will lie at much larger momentum
fractions than its first moment (\ref{eq3.2}). To be specific: in an $E=100$ GeV jets,
measured at LEP2 and initiated by light quarks (up, down, strange), the leading 
hadron typically carries a momentum fraction of $\langle z \rangle_{jet} \sim 1/4$.
In contrast, leading hadrons tend to carry a momentum fraction of the order of 
$\langle z \rangle_{lead.hadr.} \sim 3/4$.
The difference in these average $z$-values indicates the numerical importance of
trigger biases.

To sum up: if one requires single hadrons with momentum $p_T^h$  (trigger condition),
then the yield of such hadrons will be dominated by biased fragmentation processes
in which the parent parton looses as little energy as possible into the production of
subleading hadrons. This is so, since high-$p_T$ partons are rare, and the bias on
the fragmentation pattern will be more severe for larger values of $n$, when high-$p_T$
partons are rarer.

 [We note as an aside that while equations (\ref{eq3.1}) and (\ref{eq3.2}) are useful for 
 illustrating the above argument, they remain schematic since we do not specify here the 
kinematical boundaries of the $z$-integration, and we do not specify the dependence of 
the fragmentation functions on the evolution scale, which may be different for  
(\ref{eq3.1}) and (\ref{eq3.2}).] 

\noindent
{\bf Trigger biases in the medium.} Let us assume that a highly energetic parton, produced
in a hard partonic process $ \frac{d\sigma^{p+p\to f+X}}{dp_T} $, looses with 
probability $P(\epsilon)$ a fraction $\epsilon$ of its initial momentum due to
the presence of the medium, prior to hadronizing outside the medium. The 
average medium-induced fractional energy loss of such partons is
\begin{equation}
	\left(\frac{\Delta E}{E}\right)_{\rm average} = \int d\epsilon\, \epsilon\, P(\epsilon)\, .
	\label{eq3.3}
\end{equation}
However, in close analogy to the discussion of trigger biases in the vacuum,
the medium-modification of the partonic cross section will be~\cite{Baier:2001yt}
\begin{equation}
	\frac{d\sigma^{\rm med}(p_T)}{dp_T^2}
	= \int d\epsilon\, P(\epsilon)\, \frac{d\sigma(p_T/\epsilon)}{dp_T^2}
	\label{eq3.4}
\end{equation}
If the partonic cross sections falls like $\propto 1/p_T^n$, then the
medium-modified partonic cross section will be sensitive to the
typical energy loss 
\begin{equation}
	\left(\frac{\Delta E}{E}\right)_{\rm typical} = \int d\epsilon\, \epsilon^n\, P(\epsilon)\, .
	\label{eq3.5}
\end{equation}
The typical energy lost by the parent parton of a triggered high-$p_T^h$
hadron is much smaller than the average energy lost by parent partons of the same
momentum. For a partonic cross section which is steeply falling in $p_T$, the measured
high-$p_T^h$ hadrons are those which got away with the least medium-induced
energy loss. 

The probability $P(\epsilon)$ of medium-induced parton energy loss
depends on properties of the medium. In particular, it increases with increasing
in-medium path length. We recall that in heavy ion collisions, there will always be an 
outer region of the collision, from which particles can propagate into the vacuum after 
little or no interaction with the medium. Since single-hadron spectra are dominated
by hadrons which emerge with less than average medium-induced energy loss,
this implies that the hadrons selected exprimentally by measuring a single-inclusive
cross section, were produced predominantly in the outer parts of the collision region. 
The measured spectrum has a  {\bf surface bias}~\cite{Muller:2002fa}. This may limit the ability of
testing the medium with single inclusive hadron spectra~\cite{Eskola:2004cr}, simply since the 
sample of hard processes selected by measuring single inclusive hadrons 
cannot be embedded deeply into the medium, and since the surface bias
induces additional uncertainties in analyzing the medium modification. 

There are several proposals to bypass this surface bias and to characterize 
the medium-modification of hard processes which were deeply embedded in the
medium. In principle, jet measurements, based on the calorimetric measurement
of jet energy flow within the high-multiplicity environment of a heavy ion collision, 
are independent of the surface bias (and other trigger biases), since the energy of 
the initial parton is conserved throughout the medium-modified parton evolution. 
Another promising approach to bypass 
surface bias effects is the study of the recoil measured in the direction opposite
to a high-$p_T$ trigger particle.  In this case, the fragmentation pattern on the 
side of the trigger particle is biased, of course, but the recoil distribution can be
expected to show little bias. By now, several studies address the question to what
extent triggering on photons, $Z$-bosons or high-$p_T$ hadrons provides
sufficiently accurate calorimetric information about the recoil.

The surface bias is by far the best know, but possibly not the only qualitatively
novel trigger bias, which arises in the high energy collision of heavy nuclei.
For instance, one often expects that in nucleus-nucleus collisions the
incoming partons can pick up transverse momentum by multiple scattering
prior to entering a hard process. In this case, triggering on a single high-$p_T$
hadron will preferably select processes for which the center of mass of the
hard interaction moves in the direction of the triggered particle. While this
effect should be small at sufficiently high transverse momentum, it may play
an important role in the region of the Cronin peak. 

%%%%%%%%%%%%%%%%%%%%%%%%%%%%%%%%%%%%%%%%%%
%
%
\boldmath
\subsection{The nuclear modification factor}
\label{sec3.2}
\unboldmath
In the absence of medium effects, the high-$p_T$ particle yield grows proportional
to the number of hard partonic interactions, which is proportional to the number 
of nucleon-nucleon collisions, 
\begin{equation}
	\frac{dN^{A\, B \to h}}{d^2p_T\, dy} = \langle N_{\rm coll}^{AB}\rangle\, 
	\frac{dN^{p\, p \to h}}{d^2p_T\, dy}\, ,\qquad \hbox{without medium effects.}
	\label{eq3.6}
\end{equation}
Here, the average number $\langle N_{\rm coll}^{AB}\rangle$ of equivalent 
nucleon-nucleon collisions in a collision between nuclei $A$ and $B$ 
is determined by a Glauber model calculation. The single 
inclusive spectrum $dN/d^2p_T\, dy$ in a nucleon-nucleon collision is determined either 
experimentally (e.g. in p+p collisions at RHIC or LHC), or theoretically within the 
framework of perturbative factorization. To characterize deviations from this benchmark, 
one introduces the nuclear modification factor
\begin{equation}
R^h_{AB}(p_T,y,{\rm centrality})= {  {{\rm d}N^{AB\to h}_{\rm medium}\over 
{\rm d}p_T\, {\rm d}y} \over
\langle N^{AB}_{\rm coll}\rangle {{\rm d}N^{pp\to h}_{\rm vacuum}\over 
{\rm d}p_T\,{\rm d}y}}\, .
\label{eq3.7}
\end{equation}
This nuclear modification factor characterizes the medium-modification of single
inclusive spectra fully. By construction, it equals unity in the absence of medium-effects, and it
decreases if the medium suppresses the production of hard particles. 
%
%
%
%%%%%%%%%%%%%%%%%%%%%%%%%%%%%%%%%%%%%%%%%%%
\begin{figure}
\vspace{0.10in}
\centerline{
\epsfysize=6.3truecm
\epsffile{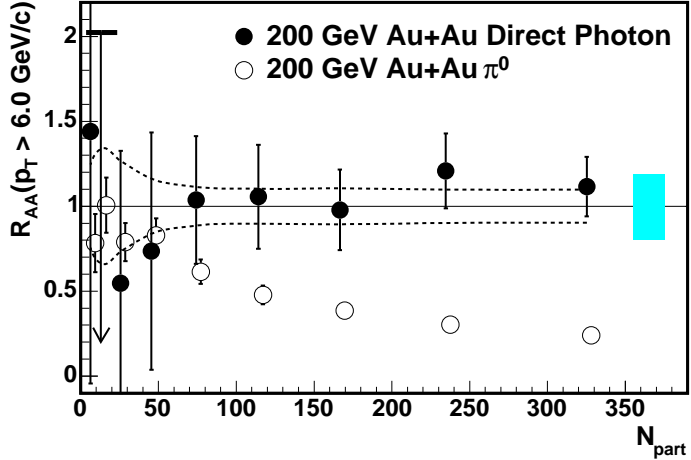}}
\caption{The nuclear modification factor (\ref{eq3.2}) as a function of centrality given
by the number of participants $N_{\rm part}$ for direct photons and neutral pions,
measured in $\sqrt{s_{\rm NN}} = 200$ GeV hadronic collisions at RHIC.
Particle yields are integrated above $p_T \geq 6$ GeV. The p+p direct photon yield
is taken from a next-to-leading order pQCD calculation with scale uncertainty indicated by
the shaded bar on the right. Dashed lines indicate the error in determining 
$\langle N^{AB}_{\rm coll} \rangle$ in (\ref{eq3.2}). All other errors are included in the
error bars. Figure taken from Ref.~\cite{Adler:2005ig}.}
\label{figraa}
\end{figure}
%%%%%%%%%%%%%%%%%%%%%%%%%%%%%%%%%%%%%%%%%%%

\noindent
{\bf The nuclear modification factor at RHIC}
Fig.~\ref{figraa} shows data for the nuclear modification factor $R_{\rm AA}$ at 
RHIC. With increasing centrality, the high-$p_T$ yield of neutral pions decreases significantly 
in comparison to the benchmark expectation (\ref{eq3.1}). 
For the most central collisions, this suppression is approximately 5-fold. In contrast,
high-$p_T$ photons appear to be unaffected within errors.  This is consistent with the
picture that the strong medium-induced suppression of high-$p_T$ hadrons is a 
final state effect, which does not occur for photons since these do not interact hadronically. 
Moreover, if one assumes that high-$p_T$ photon spectra remain unmodified, then the nuclear modification factor for photons becomes a test of the assumption that hard processes in 
heavy ion collisions scale with the number of binary nucleon-nucleon collisions, which
can be determined via a Glauber calculation of $\langle N^{AB}_{\rm coll}\rangle$.

Fig.~\ref{figraa} is but one manifestation of a generic phenomenon. In heavy ion collisions
at RHIC, {\it all} single inclusive hadron spectra are suppressed by comparable large
suppression factors. In particular, one 
observes~\cite{Arsene:2004fa,Adcox:2004mh,Back:2004je,Adams:2005dq}
:
\begin{itemize}
	\item {\it Strong and apparently $p_T$-independent suppression of $R_{AA}$ at high $p_T$~.}\\
	In $\sqrt{s_{_{NN}}}= 200$ GeV, 5-10\% central Au-Au collisions at mid-rapidity, one observes a 
	suppression of high-$p_T$ single inclusive hadron yields by a factor $\sim 5$, corresponding to 
	$R^h_{AuAu}(p_T) \simeq 0.2$ for $p_T \geq 5-7$~GeV/$c$. Within experimental errors, this 
	suppression is $p_T$-independent for higher transverse momenta in all centrality bins.
	\item {\it Evidence for final state effect.}\\
	For the most peripheral centrality bin, the nuclear modification factors measured
	at RHIC are consistent with the absence of medium-effects in both nucleus-nucleus
	($R_{AA} \sim 1$) and deuterium-nucleus ($R_{\rm dAu} \sim 1$) 
     	collisions. With increasing centrality, $R_{AA}$ decreases monotonically. In contrast,
	 no such suppression is seen in d-Au collisions. These and other observations indicate,
	 that the suppression occurs on the level of the produced outgoing partons or hadrons,
	 that it increases with increasing in-medium pathlength in the final state, and that it is 
	 hence absent in
	 d-Au collisions, where the in-medium pathlength is negligible.
	\item {\it Independence of $R_{AA}$ on hadron identity.}\\
	For transverse momenta $p_T\geq 5-7$~GeV/$c$, all identified hadron spectra
	show a quantitatively comparable degree of suppression. There is no particle-species 
	dependence of the suppression pattern at high $p_T$. Since cross
	sections for different hadron species differ widely, the species-independence of
	high-$p_T$ $R_{AA}$ indicates that the mechanism responsible for suppression
	occurs prior to hadronization.
     \end{itemize}
We emphasize that the suppression of $R_{AA}$ for hadrons is one of the
strongest medium-modifications observed in heavy ion collisions at RHIC, and that it is
a generic phenomenon found in all high-$p_T$ hadron spectra and persisting up to
the highest transverse momenta or $O(20)$ GeV measured at RHIC. The strength and
$p_T$-independence of this phenomenon supports the 
view that high-$p_T$ hadron suppression will persist at LHC to much higher transverse
momentum. Moreover, the above observations suggest to base a dynamic understanding 
of high-$p_T$ hadron suppression
on the medium-induced energy loss of high energy final state partons prior to hadron
formation. As a consequence, the standard modeling of single inclusive hadron spectra
proceeds by supplementing a pQCD factorized formalism (\ref{eq2.1}) for single inclusive spectra
with a medium modification of the produced partons prior to hadronization in the final state,
schematically
\begin{equation}
   d\sigma^{AA\to h+X}_{(\rm med)}
      =\sum_f d\sigma^{AA\to f+X}_{\rm (vac)}
       \otimes P_f(\Delta E,L,\hat q)
       \otimes D^{\rm (vac)}_{f\to h}(z,\mu^2_F)\, .
      \label{eq3.8}
\end{equation}
Here, the model-dependent input is in 
the probability $P_f(\Delta E,L,\hat q)$ of loosing a fraction $\Delta E$ of the parton 
energy while propagating over a path-length $L$ inside a medium characterized
by properties such as the quenching parameter $\hat{q}$. Experimental data from
RHIC are reproduced in models, which calculate $P_f(\Delta E,L,\hat q)$ for
mechanisms of medium-induced radiative energy loss and which take the 
nuclear geometry properly into account~\cite{Wang:2003aw,Eskola:2004cr,Dainese:2004te,Renk:2007mv,Bass:2008rv}
Some models also identify a non-negligible
role for parton energy loss via elastic interactions~\cite{Wicks:2005gt}. 
For a review of comparisons of these models to RHIC data, see 
Ref.~\cite{d'Enterria:2009am,TECHQM}.
%
%%%%%%%%%%%%%%%%%%%%%%%%%%%%%%%%%%%%%%%%%%%%
\subsection{Triggered two-particle correlations}
\label{sec3.3}
Triggered two-particle correlations are measurements, in which a trigger hadron of
high transverse momentum $p_T^{\rm trig}$ is correlated with an associated hadron
of momentum $p_T^{\rm assoc}$. As a function of azimuthal angle $\Delta \phi$, there
are two qualitatively distinct classes of measurements:

\noindent
{\bf Triggered near-side two-particle correlations.}
A measurement of two sufficiently high-$p_T$ hadrons, which are close in $\Delta \eta$ 
and $\Delta \phi$ characterizes two particles from the fragmentation pattern of the same
parent parton (assuming that there are no issues with background subtraction).
However, triggered hadrons select highly biased fragmentation patterns, in which
$p_T^{\rm trig}$ carries $O(3/4)$-th of the total energy of the parent parton, see 
section~\ref{sec4.1}. As a consequence, also the distribution of associated hadrons
is strongly biased. 

In Au-Au collisions at RHIC, the yield of high-$p_T$-trigger particles decreases by
a factor 5 from peripheral to central collisions, in accordance with the measured
nuclear modification factor. However, the yield, $\Delta \phi$-width and charge
correlation in triggered near-side two-particle correlations is insensitive to the centrality of Au-Au
collisions and coincides with the measurement in d-Au collisions~\cite{Adler:2002tq,Adams:2006yt}. 
These data are consistent with the picture of an extreme surface bias, according to
which high-$p_T$ triggers select those parent partons, whose fragmentation pattern
is unmodified by the medium. 

We are not aware of any study to what extent this surface bias could be overcome
at the LHC by triggering on significantly higher transverse momenta than the high
threshold trigger $8\, {\rm GeV} < p_T^{\rm trig} < 15\, {\rm GeV}$ used at RHIC.
However, since a high-$p_T$ trigger on a steeply falling distribution will always 
select those particles which got away with the least energy loss, it is conceivable
that the qualitative features observed at RHIC remain unchanged. At present, there
is no compelling argument that triggered near-side two-particle correlations 
are sensitive to medium effects which are not yet characterized by the nuclear
modification factor. 

\noindent
{\bf Triggered away-side two-particle correlations.}
Such measurements are expected to characterize correlations between the most
energetic leading hadron of a parent parton, and the most energetic hadronic
fragment of the recoiling parent parton. The triggered hadron will be associated
to a strongly trigger-biased fragmentation pattern. However, if one does not
apply cuts to $p_T^{\rm assoc}$, the recoiling particles may be expected to
emerge from an unbiased fragmentation pattern of the recoil parton. 

For intermediate $p_T$ triggers ($4\, {\rm GeV} < p_T^{\rm trig} < 6\, {\rm GeV}$,
data for Au-Au collisions at RHIC show that the associated particle yield for
$p_T^{\rm assoc} > 2$ GeV disappears as a function of centrality. If the 
trigger threshold is raised to higher values ($8\, {\rm GeV} < p_T^{\rm trig} < 15\, {\rm GeV}$),
the backside particle reappears again, but with strongly decreased yield. 
On a qualitative level, this demonstrates that by triggering on a high-$p_T$
particle, one can embed the recoil in such a way, that it is sensitive to the medium.
On the other hand, the azimuthal distribution of recoiling hadrons associated
to high-$p_T$ triggers does not show characteristic medium-effects, such as
broadening~\cite{Adams:2006yt}. 
This may be due to the fact that azimuthal broadening is very weak.
However, it might also indicate a surface bias, according to which 
the medium effect on the recoil either vanishes, or acts so strongly that it removes the event from
the two-particle correlation.  It has been suggested that such an effect may result
from a surface bias, which selects predominantly back-to-back particle production 
tangential to the collision region~\cite{Loizides:2006cs}.

Aside of Monte Carlo event generators, which are tailored to the simulation of
muli-particle final states, there is also one recent analytical approach towards
calculating triggered two-particle correlation functions and their medium
dependence~\cite{Majumder:2004wh,Majumder:2004br}.

We finally mention a third class of correlation measurements:\\
{\bf Triggered away-side particle-jet (photon/Z-boson)-jet correlations.}
Ideally, if one triggers on a prompt photon or $Z$-boson, one knows the energy of
the recoil parton. For sufficiently high trigger $p_T$, the recoil will be a jet, 
whose medium-modification can be characterized above background 
by all the quantities discussed in section~\ref{sec2}. Since photons and $Z$-bosons do not 
interact with the medium, one expects that the hard vertex at which the jet is
produced, is distributed homogeneously over the transverse plane (no surface bias).
Such measurements have been advocated~\cite{Wang:1996yh} as an alternative to unbiased calorimetric
jet determinations. In practice, one must understand to what extent the triggered
photons are prompt. The need to reject efficiently photons from $\pi^0$-decays, and
the need to accumulate sufficient yield limits such correlation measurements
to the range of $p_T^{\rm trig} < 70$ GeV at the LHC~\cite{Arleo:2004xj,Arleo:2006xb}. 
$Z$-boson triggered jet samples are free from this difficulty but may be statistics limited. 
To bypass the constraints of limited statistics, there are also
ideas that the fragmentation patterns of high-$p_T$ triggered hadrons may be 
understood sufficiently well to obtain useful characterizations of the recoil energy.
We expect that such ideas will be scrutinized further in the coming years in a tight interplay
between theory and experiment. 

%%%%%%%%%%%%%%%%%%%%%%%%%%%%%%%%%%%%%%%%%
\subsection{Features in the underlying event associated to high-$p_T$ triggers}
\label{sec3.4}
By triggering on a high-$p_T$ particle or calorimetric tower, one selects events
which differ from minimum bias also with respect to their soft particle distribution.
Some, but not necessarily all of these differences may arise from the soft
fragments of the high-$p_T$ final state partons. One may group conceivable
effects into two classes:

\noindent
{\bf Medium-modifications of the parton fragmentation at large $\xi$, i.e. at low $p_T$.}
Jet quenching implies an increase of soft particle multiplicity above background, which may 
be visible above background, see section~\ref{sec2.3.3}. A number of medium effects
have been suggested, which may leave characteristic imprints on this soft
yield associated to high-$p_T$ triggers. For instance, medium-induced gluon radiation
may result in the broadening of the soft multiplicity in a jet~\cite{Salgado:2003rv}, and it 
may give rise to a double-peaked structure in two-particle correlations~\cite{Polosa:2006hb}.
Also, the soft multiplicity in a jet may be distorted characteristically by a flow 
field~\cite{Armesto:2004pt}, or it may receive additional contributions from 
target components, which are kicked by elastic recoil effects into the jet cone.
Possibly the most characteristic modification would be the emergence of a Mach cone,
resulting from the fact that the parton projectile looses energy by exciting
sound modes in the medium~\cite{CasalderreySolana:2006sq,CasalderreySolana:2004qm}.
One common hallmark of all these effects is that  the energy contained
in the soft structure above background and the energy contained in the high-$p_T$
trigger must add up to the energy of the parent parton, which in principle can be constrained
independently. 

\noindent
{\bf Features in the underlying event not related to final state parton fragmentation.}\\
In a hadronic collision which contains a jet, not all the soft hadron multiplicity
produced above background results from the fragmentation of the outgoing hard
parent partons. It is known since long that requiring a high transverse
momentum hadronic structure in the collision increases the multiplicity in the 
underlying event over a wide range in rapidity~\cite{Arnison:1983gw}. At least part of 
this so-called pedestal
of the underlying event results from the fact that triggering on a high-$Q^2$ process
implies additional initial state radiation, which is spread in rapidity. 
 
For the study of heavy ion collisions, this effect may be interesting, since it sugggests
a way of adding in a controlled and localized way additional multiplicity and energy
to an event in a region which is well-separated from the high-$p_T$ trigger. One
may then be in a position to study how the medium interacts with this perturbation. 
For instance, if the pedestal is carried by a transverse flow field, then it could be moved 
predominantly in the azimuthal direction of the triggered hadron since this direction is
selected predominantly by the surface bias. A hadronic scenario closely related to
this picture has been discussed in Ref.~\cite{Voloshin:2003ud}.
 The result would be a "ridge", that is, an 
enhancement of multiplicity in a small $p_T^{\rm assoc}$-range but over 
a wide range of rapidity, which shows up only at the side of the high-$p_T$ trigger. 
There is some evidence for such a structure at RHIC. LHC may help to clarify the
dynamical origin of such structures since with increasing $\sqrt{s}$, the hadronic
activity in both the incoming and outgoing state is expected to increase significantly
with the trigger $p_T$ and may manifest itself in a wider range of $p_T^{\rm assoc}$. 

%%%%%%%%%%%%%%%%%%%%%%%%%%%%%%%%%%%%%%%%%%%%%%
%
%
%
\boldmath
\section{High-energy parton propagation in dense QCD matter}
\label{sec4}
\setcounter{equation}{0}
\unboldmath
At high transverse energy the medium-modification of jets and of single inclusive 
hadron spectra and jet-like hadron correlations is dominated by modification of the 
partonic propagation prior to hadronization. This is supported by data from RHIC 
(see subsection~\ref{sec3.1}) and by parametric 
estimates (see subsection~\ref{sec2.1}). It motivates us in the present chapter
to review aspects of the in-medium propagation of partons. 

The interaction of an energetic projectile with a component of a medium can be 
classified according to whether it proceeds via elastic processes (multiple scattering
without particle production) or via inelastic processes. In the phenomenology of
heavy ion collisions, these processes are often referred to as radiative and collisional
parton energy loss, respectively. In addition, virtual partons can split without interacting
with the medium. In the following, we discuss these three components of in-medium
parton propagation. We start by discussing the ultra-relativistic limit of very large
projectile energy ($E \to \infty$), where parton propagation can be described in 
the eikonal formalism. 

%%%%%%%%%%%%%%%%%%%%%%%%%%%%%%%%%%%%%%%%%%%%%
\subsection{Parton propagation through dense QCD matter in the eikonal formalism}
\label{sec4.1}
In-medium parton propagation can be described by solving the Dirac equation for
the wave function of a parton in the spatially extended color field of the target. 
In the limit of infinite projectile energy, the solution of this Dirac equation can
be written in terms of the 
eikonal factor
\begin{equation}
  W({\bf x}_i)={\cal P}\exp\{i\int dz^-T^aA^+_a({\bf x}_i, z^-)\}\, .
  \label{eq4.1}
\end{equation}
Here, $A^+$ is the large component of the target color field and $T^a$ is the generator of 
$SU(N)$ in the representation corresponding to a given parton. Here and in what follows, 
we use bold-face variables such as ${\bf x}_i$ to denote 2-dimensional vectors which lie
in the plane orthogonal to the beam direction. Equation (\ref{eq4.1}) 
is the specific form of the phase factor
in the light cone gauge $A^-= 0$ for a projectile moving  in the negative $z$ direction, so
that the light cone coordinate $x^+ = (z+t)/\sqrt{2}$ does not change during propagation
through the target. The phase factor takes a different form in other gauges or other
Lorentz frames, but the final result is gauge invariant and Lorentz covariant, of course.

To be specific, let us consider a high-energy quark of color $\alpha$ at transverse position 
${\bf x}$ with incoming wave function $\vert \alpha({\bf x})\rangle$. Scattering on the target
results in the outgoing wave function 
\begin{equation}
 {\cal S} \vert \alpha({\bf x})\rangle =  W({\bf x})_{\beta\alpha} \vert \alpha({\bf x})\rangle\, .
 \label{eq4.2}
 \end{equation}
 This wave function is the solution of the Dirac equation in the external color field
 $A^+$ and in the limit of infinite projectile energy. More generally,  the eikonal
 formalism describes the interaction of any set of  projectile partons with the target by
a color rotation $\alpha_i \to \beta_i$ of each projectile
component $i$, resulting in an eikonal phase $W({\bf x}_i)_{\alpha_i \beta_i}$. 
The out-going wave function $W({\bf x})_{\beta\alpha} \vert\{\alpha,{\bf x}\}\rangle$. 
Here, the physics is that in the ultra-relativistic limit, the target is Lorentz contracted to an
infinitesimally thin pancake In the rest frame of the projectile. As a consequence, the 
projectile cannot change its transverse position during the interaction, it propagates 
on an eikonal straight-line trajectory. In this limit, interactions with the target are
recoilless (and hence, there is no collisional energy loss) and they lead to the 
color rotation of each projectile component according to (\ref{eq4.1}). If 
different components acquire a different relative phase due to interactions with 
the target, then these components of the incoming hadron wave function decohere
in the scattering. This decoherence is at the basis of many formulations of radiative
energy loss, and hence we discuss it here explicitly in the simplest case within
the eikonal formalism. For more details about the eikonal formalism, we 
refer to Ref.~\cite{Kovner:2001vi} and references therein.
 
 \noindent
{\bf Example: gluon radiation off high-energy quark in the eikonal formalism.}
We consider a high energy quark, which impinges on the target with a fully developed
wave function. In the first order in perturbation theory the incoming 
wave function contains the Fock state $ \vert \alpha\rangle$ 
of the bare quark, supplemented by the coherent state of quasi real 
gluons which build up the Weizs\"acker-Williams field $f({\bf x})$, 
\begin{eqnarray}
  \vert \Psi_{\rm in}^\alpha\rangle &=& 
  \vert \alpha\rangle + \int d{\bf x}\,d\xi f({\bf x})\, 
  T^b_{\alpha\, \beta}\, \vert \beta\, ; b({\bf x},\xi)\rangle\, .
  \label{eq4.3}
\end{eqnarray}
Here Lorentz and spin indices are suppressed. In the projectile 
light cone gauge $A^-=0$, the gluon field of the projectile is 
the Weizs\"acker-Williams field
\begin{equation}
  A^i({\bf x})\propto  \theta(x^-)\, f_i({\bf x})\, ,\qquad \qquad
  f_i({\bf x})\propto g{{\bf x}_i\over {\bf x}^2}\, ,
  \label{eq4.4}
\end{equation}
where $x^-=0$ is the light cone coordinate of the quark in the wave 
function. The integration over the rapidity of the gluon in the wave 
function (\ref{eq4.3}) goes over the gluon rapidities smaller than that 
of the quark. In the leading logarithmic order the wave function 
does not depend on rapidity and we suppress the rapidity label in the following. 
The outgoing wave function reads
\begin{eqnarray}
 \vert \Psi_{\rm out}^\alpha \rangle &=& 
  W^F_{\alpha\, \gamma}({\bf 0})\, \vert\gamma\rangle +
  \int   d{\bf x}\, f({\bf x})\, T_{\alpha\, \beta}^b
  W_{\beta\, \gamma}^F({\bf 0})\, W_{b\, c}^A({\bf x})\, 
  \vert \gamma\, ;c({\bf x})\rangle\, ,
  \label{eq4.5}
\end{eqnarray}
where  $W^F({\bf 0})$ and $W^A({\bf x})$ are the Wilson lines in the 
fundamental and adjoint representations respectively, corresponding 
to the propagating quark at the transverse position ${\bf x}_q={\bf 0}$ 
and gluon at ${\bf x}_g={\bf x}$. 

The projection of the outgoing wave function $\Psi_{\rm out}$ on the
subspace spanned by incoming quark wave functions will describe
outgoing quarks which are 'dressed' with gluons. Hence, 
to count the number of newly produced gluons in the state (\ref{eq4.5}),
one must select the projection of $\Psi_{\rm out}$ on the subspace
orthogonal to the incoming states
\begin{eqnarray}
  && \vert \delta \Psi_\alpha \rangle =
  \vert \Psi_{\rm out}^\alpha\rangle -
  \sum_\gamma    \vert \Psi_{\rm in}(\gamma)\rangle \langle \Psi_{in}(\gamma) 
   \vert \Psi_{out}^\alpha\rangle\, .
  \label{eq4.6}
\end{eqnarray}
Here, the index $\gamma$ in the projection operator runs over  
the quark color index, so that the second term in (\ref{eq4.6}) projects
out the entire Hilbert subspace of incoming states. 
 
The number spectrum of produced gluons is obtained by calculating 
the number of gluons in the state $\delta \Psi_\alpha$, averaged over the incoming color index 
$\alpha$. After some color algebra, one obtains
\begin{eqnarray}
  && N_{\rm prod}({\bf k}) = \frac{1}{N} \sum_\alpha
  \langle \delta \Psi_\alpha\vert a_d^\dagger({\bf k})\, 
  a_d({\bf k})\vert\, \delta\Psi_\alpha \rangle
  \nonumber\\
  && \quad 
  = \frac{\alpha_s\, C_F}{2\pi}\, 
  \int d{\bf x}\, d{\bf y}\, e^{i{\bf k}\cdot({\bf x}-{\bf y})}
  \frac{ {\bf x}\cdot {\bf y}}{ {\bf x}^2\, {\bf y}^2}
%  f({\bf x})\, f({\bf y})
  \Bigg[ 1 - \frac{1}{N^2 - 1}\, 
  \langle\langle
  {\rm Tr}\left[ W^{A\, \dagger}({\bf x})\, W^A({\bf 0}) \right]
  \rangle\rangle_t
  \nonumber\\
&& \qquad \qquad \qquad \qquad \qquad \qquad  - \frac{1}{N^2 - 1}\, 
  \langle\langle
  {\rm Tr}\left[ W^{A\, \dagger}({\bf y})\, W^A({\bf 0}) \right]
  \rangle\rangle_t
  \nonumber \\
&& \qquad \qquad \qquad \qquad \qquad \qquad
  +\frac{1}{N^2 - 1}\, 
  \langle\langle
  {\rm Tr}\left[ W^{A\, \dagger}({\bf y})\, W^A({\bf x}) \right]
  \rangle\rangle_t\Bigg]\, .
  \label{eq4.7}
\end{eqnarray}
Here, we have used $f({\bf x})\, f({\bf y}) = \frac{\alpha_s}{2\pi} \, 
\frac{{\bf x}\cdot {\bf y}}{{\bf x}^2\, {\bf y}^2}$ for the
Weizs\"acker-Williams field of the quark projectile in configuration 
space and the symbol $\langle\langle\dots\rangle\rangle_t$ denotes the averaging 
over the gluon fields of the target.

We emphasize three findings, which are seen explicitly in the eikonal formalism
presented here, and which persist in a more general treatment:
\begin{itemize}
\item {\it Radiative energy loss dominates over collisional energy loss at high energy.}\\
At high projectile energy, momentum transfer between projectile and target
is predominantly transverse. Longitudinal momentum transfer is suppressed by
powers of projectile energy. Hence, in the eikonal formalism the projectile does not 
transfer longitudinal momentum to target components via elastic interactions. In
contrast, the eikonal formalism allows for the calculation of inelastic processes,
as discussed above. This illustrates that radiative mechanisms are the dominant
source of medium modification for sufficiently high parton energy. 
\item {\it A light-like Wilson loop defines the medium-dependence of the gluon spectrum. }\\
As seen from the radiation spectrum (\ref{eq4.7}), the entire information 
about the target resides in the target average of two light-like adjoint Wilson lines.
Although the presence of quarks leads to the appearance of fundamental Wilson 
lines in intermediate stages of the calculation, see e.g. equation (\ref{eq4.5}), the 
averaging involved in (\ref{eq4.7}) combines them into adjoint ones with the help
of the Fierz identity 
$W_{ab}^F({\bf x}) = 2\, {\rm Tr}\left[T^a\, W^{F\dagger}({\bf x})T^b \, W^{F}({\bf x}) \right]$.
\item {\it Particle production in the eikonal formalism is determined by a decoherence  effect.}\\
The above calculation provides an explicit example for the general statement, that
gluons are emitted from a parton projectile if and only if they have accumulated due
to medium interactions a relative phase of order unity with respect to other partonic
projectile components. 
\end{itemize}
%

%%%%%%%%%%%%%%%%%%%%%%%%%%%%%%%%%%%%%%%%%%
\subsection{Gluon radiation off quarks produced in the medium}
\label{sec4.2}
In section~\ref{sec4.1}, we considered radiation off a quark, which propagated a long
distance before impinging on the target. This quark had a fully evolved wave function, that
means, it had time to develop a gluon cloud around it. Medium-induced radiation
amounts to the partial stripping of these quasi real gluons in the quark wavefunction.
In the present section, we discuss medium-induced gluon radiation off a 
parton, which is produced in a large momentum transfer process {\it inside the medium}. 
This problem is significantly more complicated mainly 
because of two issues:
\begin{itemize}
\item {\it Interference between radiation in the vacuum and medium-induced radiation}\\
In the absence of a medium, a parton produced in a hard process will radiate its large 
virtuality $Q$ on a typical timescale $1/Q$ by developing a parton shower. In 
the rest frame of the medium, this time scale is Lorentz dilated by a factor $E_{\rm parton}/M$, 
where the parton mass is $M \sim Q$. Typical radiation times $\sim E_{\rm parton}/Q^2$
are comparable to the typical in-medium pathlengths in a nucleus-nucleus collision. As
a consequence, one expects an interference pattern between the radiation present in
the vacuum, and the additional radiation induced due to scattering in the medium.
\item {\it Corrections to eikonal approximation}\\
In the ultra-high energy (eikonal) approximation, the longitudinal extension of the target
is contracted to a delta function. As a consequence, gluon radiation off a hard parton
occurs either before or after the target, but not within the target.  In contrast, to
take interference effects into account, it is important to locate the gluon emission vertex
inside the medium. This requires a formulation which is sensitive to longitudinal distances
(or the time spent) in the medium. The momentum conjugate to this distance is the 
projectile energy $E$ or light cone energy 
$p_-$. So, to place an emission vertex within the medium, one has to keep track 
at least of the $O(1/p^-)$-corrections to the eikonal formalism.
\end{itemize}
%
%%%%%%%%%%%%%%%%%%%%%%%%%%%%%%%%%%%%%%%%%%%%
\subsubsection{Gluon radiation in the path integral formalism}
\label{sec4.2.1}
In the following, we present the main elements of a formulation which goes beyond the 
eikonal approximation and accounts for interference effects between vacuum and 
medium-induced radiation. Up to order $O(1/p^-)$, one can write the solution 
$\Psi_{\rm out}({\bf r}_{\rm out},x_{\rm out}^-) = \int d{\bf r}_{\rm in}\, 
G({\bf r}_{\rm in},x_{\rm in}^-;{\bf r}_{\rm out},x_{\rm out}^-\vert p^-)\, 
\Psi_{\rm out}({\bf r}_{\rm in},x_{\rm in}^-)$ of the 
Dirac equation for a colored partonic projectile propagating in a spatially extended 
color field $A^+$ in terms of the light-cone Green's 
function~\cite{Kopeliovich:1998nw,Zakharov:1996fv,Zakharov:1997uu,Zakharov:1998sv}
\begin{eqnarray}
	&&G({\bf r}_{\rm in},x_{\rm in}^-;{\bf r}_{\rm out},x_{\rm out}^-\vert p^-)
		= \hspace{-0.5cm}
		\int \limits_{{\bf r}(x_{\rm in}^-) = {\bf r}_{\rm in}}^{{\bf r}(x_{\rm out}^-) = {\bf r}_{\rm out}}
		 \hspace{-0.5cm}
			{\cal D}{\bf r}(\xi)\, \exp\left[ i \frac{p^-}{4} \int_{x_{\rm in}^-}^{x_{\rm out}^-} 
			  d\xi\, \dot{\bf r}^2(\xi) \right]\, W({\bf r}(\xi); x_{\rm in}^-},{x_{\rm out}^-)\, ,
			  \label{eq4.8}\\
          && \qquad \qquad W({\bf r}; x_{\rm in}^-},{x_{\rm out}^-) = {\cal P}\, \exp
          	\left[  i \int_{x_{\rm in}^-}^{x_{\rm ou}^-} d\xi\, A^+({\bf r}(\xi),\xi) \right]\, .
	\label{eq4.9}
\end{eqnarray}
This solution contains a non-eikonal Wilson line,
which 'wiggles' in transverse position along a path ${\bf r}(\xi)$. In the limit of ultra-high
parton energy, $p^- \to \infty$, when the finite energy corrections of order $O(1/p^-)$
vanish, this expression reduces to the eikonal Wilson line (\ref{eq4.1}), 
\begin{equation}
	\lim_{p^- \to \infty} G({\bf r}_{\rm in},x_{\rm in}^-;{\bf r}_{\rm out},x_{\rm out}^-\vert p^-)
 	= W({\bf r}_{\rm in}; x_{\rm in}^-,x_{\rm out}^-)
	\, \delta^{(2)}({\bf r}_{\rm out}-{\bf r}_{\rm in})\, .
	\label{eq4.10}
\end{equation}
In the eikonal formalism, gluon radiation (\ref{eq4.7}) is determined by 
the target average of two eikonal Wilson lines $ \langle\langle
{\rm Tr}\left[ W^{A\, \dagger}({\bf y})\, W^A({\bf 0}) \right]  \rangle\rangle_t$.
In close analogy, it is the target average of pairs of Green's functions 
${\cal K}$, which determines gluon radiation in the present formalism. 
This target average can be defined in terms of the two-point correlation
function of the target color field, see Ref.~\cite{Kovner:2003zj,Wiedemann:1999fq} 
for technical details. 
If this color field is parametrized by a set of static scattering centers of
path-dependent density $n(\xi)$, then the target average can be written as
~\cite{Zakharov:1996fv,Wiedemann:2000za}
\begin{eqnarray}
  {\cal K}\bigl({\bf r}',z';{\bf r},z|\mu\bigr)
  &=&   \int {\cal D}{\bf r}\, 
  \exp\left[  i\, \int\limits_z^{z'}\, d\xi\,
  \left[{\mu\over 2}\dot{\bf r}^2 
  + i\, \frac{1}{2} n(\xi)\, \sigma\left({\bf r}\right)
  \right] \right]\, .
  \label{eq4.11}
\end{eqnarray}
Here, $\sigma\left({\bf r}\right)$ is the so-called dipole cross section, which
is defined in terms of the Fourier transform of the elastic cross section betwee
 scattering center and target. One novel feature of the target average of two
 Green's functions (\ref{eq4.8}) is that the average depends on the energies
 $\alpha p$ and $(1-\alpha)p$ in the arguments of both Green's functions. 
For this reason, ${\cal K}$ in (\ref{eq4.11}) depends on  $\mu \equiv \alpha (1-\alpha)p$. 
In accordance with the notation used in
parton energy loss calculations, we have changed from light-cone coordinates
to the longitudinal $z$. 
One can check that the $\mu\to\infty$ limit of (\ref{eq4.11}) leads to the 
average of two eikonal Wilson lines, entering (\ref{eq4.6}).

The inclusive energy distribution of gluon radiation off an in-medium 
produced parton can be expressed in terms of the path integral (\ref{eq4.11}) as
~\cite{Wiedemann:2000za}
\begin{eqnarray}
  \omega\frac{dI}{d\omega}
  &=& {\alpha_s\,  C_R\over (2\pi)^2\, \omega^2}\,
    2{\rm Re} \int_{\xi_0}^{\infty}\hspace{-0.3cm} dy_l
  \int_{y_l}^{\infty} \hspace{-0.3cm} d\bar{y}_l\,
   \int d{\bf u}\,  \int_0^{\chi \omega}\, d{\bf k}\, 
  e^{-i{\bf k}\cdot{\bf u}}   \,
  e^{ -\frac{1}{2} \int_{\bar{y}_l}^{\infty} d\xi\, n(\xi)\,
    \sigma({\bf u}) }\,
  \nonumber \\
  && \times {\partial \over \partial {\bf y}}\cdot
  {\partial \over \partial {\bf u}}\,
  \int_{{\bf y}=0}^{{\bf u}={\bf r}(\bar{y}_l)}
  \hspace{-0.5cm} {\cal D}{\bf r}
   \exp\left[ i \int_{y_l}^{\bar{y}_l} \hspace{-0.2cm} d\xi
        \frac{\omega}{2} \left(\dot{\bf r}^2
          - \frac{n(\xi) \sigma\left({\bf r}\right)}{i\, \omega} \right)
                      \right]\, .
    \label{eq4.12}
\end{eqnarray}
Here, ${\bf k}$ denotes the transverse momentum of the emitted gluon.
The two-dimensional transverse coordinates ${\bf u}$, ${\bf y}$
and ${\bf r}$ emerge in the derivation of (\ref{eq4.12}) as distances
between the positions of projectile components in the amplitude
and complex conjugate amplitude. The longitudinal coordinates
$y_l$, $\bar{y}_l$ integrate over the ordered longitudinal
gluon emission points in amplitude and complex conjugate amplitude.
The limit $\vert{\bf k}\vert < \chi\, \omega$
on the transverse phase space restricts gluon emission to a finite opening 
angle $\Theta$, $\chi = \sin\Theta$. For the
full angular integrated quantity, $\chi = 1$. 

There are two limiting cases, in which the compact expression (\ref{eq4.12})
for the medium-induced gluon energy distribution can be related to several
currently used formalism of parton energy loss:

\noindent
{\bf Opacity expansion:}~\cite{Wiedemann:2000za,Gyulassy:2000er}
An expansion in powers of opacity is obtained by 
expanding the integrand of (\ref{eq4.12}) in powers of the dipole cross section
$\sigma({\bf r})$. In so-called Gyulassy-Wang models~\cite{Gyulassy:1993hr}
, where the color field
strength of the medium is parametrized in terms of a set of static scattering
centers of density $n(\xi)$, 
\begin{equation}
 \sigma({\bf r}) = 2 \int \frac{d{\bf q}}{(2\pi)^2}\, \vert A({\bf q})\vert^2\, 
 	\left(1 - \exp\{ i\, {\bf q}.{\bf r}\} \right)\, .
	\label{eq4.13}
 \end{equation}
 Here, $A({\bf q})$ is the potential of a single scattering center (for more details,
 including the treatment of color in potential scattering, see Ref.~\cite{Kovner:2003zj}).
 
 In the absence of a medium, that is, to zeroth order in opacity, one finds
\begin{equation}
  \omega {d^3I(N=0)\over d\omega\, d{\bf k}}
  = {\alpha_s\over \pi^2}\, 
    C_R\, \frac{1}{{\bf k}^2}\, ,
    \qquad   H({\bf k}) =  \frac{1}{{\bf k}^2}\, .
  \label{eq4.14}
\end{equation} 
This is the characteristic $1/{\bf k}^2$-gluon radiation spectrum associated to the DGLAP
parton branching process in the vacuum. Expression (\ref{eq4.14}) illustrates that the
gluon energy distribution (\ref{eq4.12}) contains
information about the off-shellness of the projectile parton, since it allows for
gluon radiation without medium interaction.

For a medium of constant 
density $n(\xi) = n_0$ and longitudinal extension $L$, the contribution to (\ref{eq4.12})
to first order of opacity takes the form
\begin{equation}
  \omega {d^3I(N=1)\over d\omega\, d{\bf k}}
  = {\alpha_s\over \pi^2}\, 
    \frac{C_R}{2\,\omega^2}\, 
    \int \frac{d{\bf q}_1}{(2\pi)^2}\,
    |a_0({\bf q}_1)|^2
    \left({\bf k}\cdot {\bf q}_1\right)
    n_0\, \tau\, \tau_1^2 \left[ \frac{L}{\tau_1} - \sin\left(\frac{L}{\tau_1}\right)\right]\, ,
  \label{eq4.15}
\end{equation} 
where
\begin{equation}
	\tau= \frac{2\omega}{{\bf k}^2}\, ,\qquad
	\tau_1 = \frac{2\omega}{\left( {\bf k}  - {\bf q}_{1}\right)^2}\, .
	\label{eq4.16}
\end{equation}
This is an explicit illustration of the formation time physics discussed in section~\ref{sec2.1}. 
The interference term $\left[ \frac{L}{\tau_1} - \sin\left(\frac{L}{\tau_1}\right)\right]$
in (\ref{eq4.15}) 
prevents gluons from being radiated if $L< \tau_1$. According to the discussion of
equation (\ref{eq2.5}), $\tau_1$ can be regarded as the formation time of the emitted gluon
prior to interaction with target. The condition $L/\tau_1 < 1$ then translates into the
simple statement that a gluon can only scatter if at the time of the scattering it is an 
independent component of projectile wave function, that means, if it is formed. 
The generalization of this observation can be used for a probabilistic implementation
of medium-induced interference effects, see section~\ref{sec4.4.3}

In general, the medium-induced gluon energy distribution (\ref{eq4.12}) receives
contributions from vacuum radiation, from the elastic scattering of vacuum radiation 
on the medium, and from additional medium-induced radiation. These three contributions
can be clearly isolated in the formal limit of a large distance between the location of 
projectile production and interaction with the medium. This
can be done in the incoherent limit in which $L\to\infty$ but $n_0 L =$ fixed.
Up to first order in opacity, one finds
\begin{eqnarray}
  &&\lim_{L\to\infty}\sum_{m=0}^{N=1}
  \omega {d^3I(m)\over d\omega\, d{\bf k}}
  = {\alpha_s\over \pi^2}\, C_F
  (1-w_1)H({\bf k}) 
  \nonumber \\
  &&\qquad + {\alpha_s\over \pi^2}\, C_F
                  \, n_0\, L\, \int
		  \frac{d{\bf q}_1}{(2\pi)^2}
		  \vert a_0({\bf q}_1)\vert^2
		  \Bigg (
                  H({\bf k} + {\bf q}_{1})
                  + R({\bf k},{\bf q}_{1}) \Bigg )\, .
  \label{eq4.17}
\end{eqnarray}
Here, $H({\bf k})$ is the hard, medium-independent radiation 
(\ref{eq4.14}) reduced by the probability $w_1$ that
one interaction of the projectile occurs in the medium. The
second term describes the hard radiation component which rescatters
once in the medium. The third term is the medium-induced Gunion-Bertsch
contribution $R({\bf k},{\bf q}_{1})$ for additional gluon radiation. 
For realistic kinematical conditions, interference terms as e.g. in (\ref{eq4.16}) 
interpolate between these simple and physically intuitive limiting cases. 

It has been pointed out that the opacity expansion may have bad convergence
properties~\cite{Arnold:2008iy}. On the other hand, comparisons of numerical results for
the $N=1$ opacity expansion and the multiple soft scattering approximation 
discussed below show that both approaches can be brought to quantitative
agreement on key parameters. 

\begin{figure}[h]\epsfxsize=9.7cm
\centerline{\epsfbox{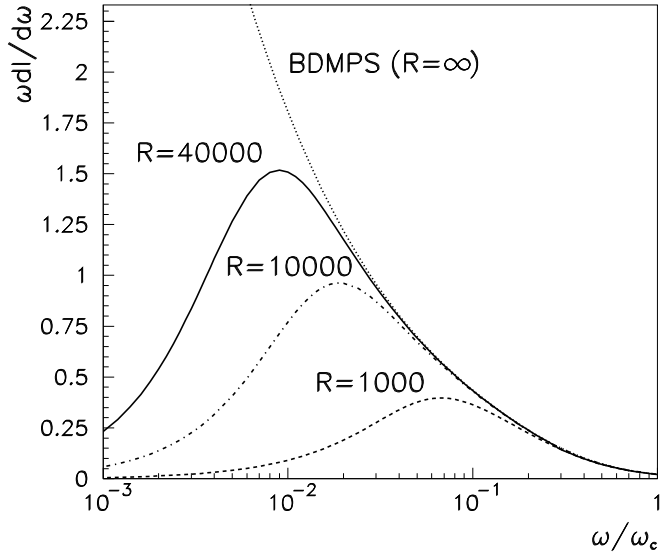}}
\vspace{0.5cm}
\caption{The medium-induced gluon energy distribution 
$\omega \frac{dI}{d\omega}$ as a function of the gluon energy $\omega$ 
in units of $\omega_c = \frac{1}{2} \hat{q}\, L^2$, and for different values of the 
kinematic constraint $R = \omega_c\, L$.
Figure taken from Ref.~\cite{Salgado:2003gb}.
}\label{fig9}
\end{figure}

 \noindent
 {\bf Multiple soft scattering approximation} 
 \cite{Baier:1996kr,Baier:1996sk,Zakharov:1996fv,Zakharov:1997uu,Wiedemann:2000za,Wiedemann:2000tf}The medium-induced
 gluon energy distribution (\ref{eq4.12}) can be studied in the 
 saddle point approximation, where it is sensitive to the short distance
 behavior of the dipole cross section 
\begin{eqnarray}
  n(\xi)\, \sigma({\bf r}) \simeq \frac{1}{2}\, \hat{q}(\xi)\, {\bf r}^2\, .
  \label{eq4.18}
\end{eqnarray}
Here, $\hat{q}(\xi)$ is refered to as BDMPS (Baier-Dokshitzer-Mueller-Peign\'e-Schiff) 
transport coefficient.  The path integral becomes that of a harmonic oscillator and can
be calculated explicitly. Typical numerical results for the medium-induced 
gluon energy distribution (\ref{eq4.12}) are shown in 
 Figure~\ref{fig9}  for a static BDMPS transport coefficient $\hat{q} = \hat{q}(\xi)$ 
 extending over a finite in-medium pathlength $L$. 
%
%%%%%%%%%%%%%%%%%%%%%%%%%%%%%%%%%%%%%%%%%%%%%%%%%%%%%%%%%%%%%%%%%%%%
%%%%%%%%%%%%%%%%%%%%%%%%%%%%%%%%%%%%%%%%%%%%%%%%%%%%%%%%%%%%%%%%%%%
%
As explored first in~\cite{Baier:1998yf}, the path integral (\ref{eq4.12}) allows for a
saddle point approximation also for the case of time-dependent
densities and quenching parameters of the form
\begin{equation}
  \hat{q}(\xi) = \hat{q}_d \left( \frac{\xi_0}{\xi} \right)^\alpha\, .
  \label{eq4.19}
\end{equation}
Here,  $\hat{q}_d$ is the value of $\hat{q}$, taken at the 
initial plasma formation time $\xi_0$. The power $\alpha = 0$ characterizes
the static medium discussed above. The value $\alpha=1$ is obtained for 
a system with one-dimensional, boost-invariant longitudinal expansion. 
In general, one can encode with a suitable choice of $\alpha$ for the
characteristic density decrease resulting from the expansion of the collision region. 
Remarkably, the radiation spectrum 
$\omega \frac{dI}{d\omega}$ satisfies a simple scaling law which 
relates the radiation spectrum of a dynamically expanding
collision region to an equivalent static scenario. The 
linearly weighed line integral \cite{Salgado:2002cd}
\begin{equation}
  \overline{\hat{q}} = \frac{2}{L^2}\int_{\xi_0}^{\xi_0+L} d\xi\, 
  \left( \xi - \xi_0\right)\, \hat{q}(\xi) 
  \label{eq4.20}
\end{equation}
defines the transport coefficient of the equivalent static
scenario. The gluon energy distribution (\ref{eq4.12}) of an expanding scenario (\ref{eq4.19})
is approximately equal to the gluon energy distribution of a static scenario with 
$\hat{q}(\xi) =  \overline{\hat{q}}$. 
The linear weight in (\ref{eq4.32}) implies that scattering centers
which are further separated from the production point of the
hard parton are more effective in leading to partonic energy
loss. 
 
%%%%%%%%%%%%%%%%%%%%%%%%%%%%%%%%%%%%%%%%%
\subsubsection{Qualitative features of medium-induced gluon radiation}
\label{sec4.2.2}

Main features of the gluon energy distribution in Fig.~\ref{fig9} can be understood
in terms of qualitative arguments. 
We consider a gluon in the hard parton 
wave function. This gluon is emitted due to multiple scattering
if it picks up sufficient transverse momentum to decohere from the partonic
projectile. For this, the average phase $\varphi$ accumulated by
the gluon should be of order one or larger,
\begin{equation}
  \varphi = \Bigg\langle \frac{k_\perp^2}{2\omega}\, \Delta z \Bigg\rangle
  \sim \frac{\hat{q}\, L}{2\omega} L = \frac{\omega_c}{\omega}\, .
  \label{eq4.21}
\end{equation}
Thus, from a hard parton traversing a finite path length $L$ in the medium,
gluons will be emitted up to a  ``characteristic gluon frequency''
\begin{equation}
  \omega_c = \frac{1}{2}\, \hat{q}\, L^2\, .
  \label{eq4.22}
\end{equation}
For an estimate of the shape of the energy distribution, we 
consider the number $N_{\rm coh}$ of scattering centers which 
add coherently in the gluon phase (\ref{eq4.21}), 
$k_T^2 \simeq N_{\rm coh}\, \langle q_T^2\rangle_{\rm med}$. 
Based on expressions
for the coherence time of the emitted gluon, 
$t_{\rm coh} \simeq \frac{\omega}{k_T^2} \simeq 
\sqrt{\frac{\omega}{\hat{q}}}$
and $N_{\rm coh} = \frac{t_{\rm coh}}{\lambda} = 
\sqrt{\frac{\omega}{\langle q_T^2\rangle_{\rm med}\, \lambda}}$, 
one estimates for the
gluon energy spectrum per unit path length
\begin{equation}
  \omega \frac{dI}{d\omega\, dz} \simeq 
  \frac{1}{N_{\rm coh}}\, 
  \omega \frac{dI^{\rm 1\, scatt}}{d\omega\, dz} \simeq
  \frac{\alpha_s}{t_{\rm coh}}
  \simeq \alpha_s\, \sqrt{\frac{\hat{q}}{\omega}}\, .
  \label{eq4.23}
\end{equation}
This $1/\sqrt{\omega}$-energy dependence of the
medium-induced non-abelian gluon energy spectrum is
expected for sufficiently small $\omega < \omega_c$. 
This dependence is seen in Fig.~\ref{fig9} to be realized by the full expression (\ref{eq4.12}),
if one neglects (as for the above estimate) kinematical constraint in transverse phase space,
which cut-off the energy distribution in the infrared. 
For the $\omega$-integrated average parton energy loss, one finds 
from the above pocket estimates by integrating the differential distribution (\ref{eq4.23}) over the
in-medium path length $L$ and over the gluon energy $\omega$ up to $\omega_c$,

\begin{equation}
 \langle \Delta E \rangle = 
 \int_0^\infty d\omega\, 
   \omega \frac{dI}{d\omega}
 = \frac{\alpha_s C_R}{2}\, \omega_c \propto \hat{q}\, L^2\, . 
 \label{eq4.24}
\end{equation}
The same parametric dependence $\propto \hat{q}\, L^2$ can be found 
$R = \omega_c\, L \to \infty$ at $\omega_c = {\rm fixed}$~\cite{Salgado:2003gb}. 
This is the famous BDMPS-result that the average radiative energy loss grows
quadratically with in-medium path length for sufficiently small $L$. 
The pocket estimate (\ref{eq4.23}) encodes the main features of the BDMPS
result, namely the correct small-$\omega$
behavior as well as the correct dependence of the average energy loss on density
and in-medium path length. 

We finally summarize the main results following from the medium-induced 
gluon energy distribution (\ref{eq4.12}):
\begin{itemize}
	\item The average parton energy loss grows parametrically like
		\begin{equation}
			\Delta E \propto L^2\, .
			\label{eq4.25}
		\end{equation}
	\item The gluon radiation shows a characteristic hierarchical dependence on the
	color charge and mass of the parton projectile
			\begin{equation}
			\Delta E_{\rm gluon} > \Delta E_{\rm light quark}
			> \Delta E_{\rm heavy quark}\, .
			\label{eq4.26}
		\end{equation}	
		Here, the first inequality follows from the larger color charge of partons in the
		adjoint representation. The second inequality is due to the dead cone effect,
		which suppresses radiation off massive particles in the vacuum and in the
		medium~\cite{Dokshitzer:2001zm,Armesto:2003jh,Zhang:2003wk,Djordjevic:2003zk}. 
		Several studies have analyzed the extent to which the hierarchy (\ref{eq4.26})
	affects the suppression patterns of heavy-flavored single inclusive
	hadron spectra and single electron spectra, which are dominated at sufficiently high
	transverse momentum by the semi-leptonic decays of heavy-flavored hadrons.
	\cite{Armesto:2005iq,Djordjevic:2005db}. For a concise mini-review of approaches
	with focus on lower transverse momentum, see Ref.~\cite{Rapp:2008zq}.
	\item The average transverse momentum of medium-induced gluon radiation grows
	as expected for Brownian motion resulting from multiple scattering,
			\begin{equation}
			\langle k_T^2\rangle \propto \hat{q}\, L\, .
			\label{eq4.27}
		\end{equation}
		\item The formalism studied in this section is recoilless. It neglects 
		contributions to parton energy loss, that result from longitudinal momentum
		transfer to target degrees of freedom. To improve on this
		point, one requires a dynamical description of target degrees of freedom.
	\item The $\omega$-dependence of the gluon energy distribution shows a 
	characteristic steepening due to medium-effects, see e.g. the 
	$\sqrt{\omega}$-modification of (\ref{eq4.23}).
\end{itemize}

%%%%%%%%%%%%%%%%%%%%%%%%%%%%%%%%%%%%%%%%%%%%
\subsubsection{Multiple gluon emission}
\label{sec4.2.3}
The formalism discussed in this section~\ref{sec4.2} is based on calculating matrix
elements for one-gluon emission. To describe the energy degradation of a
highly energetic parton in dense QCD matter, one needs to account for the
possibility that a total energy $\Delta E$ is carried away by the emission of an 
arbitrary number of $n$ gluons. In the absence of information about the medium-dependence
of $n$-gluon emission cross sections, it has been proposed~\cite{Baier:2001yt} to
treat subsequent gluon emissions as independent. The probability distribution
$P(\Delta E)$ of losing a total energy $\Delta E$ in the emission of an arbitrary
number of gluons is then 
\begin{eqnarray}
  P(\Delta E) = \sum_{n=0}^\infty \frac{1}{n!}
  \left[ \prod_{i=1}^n \int d\omega_i \frac{dI(\omega_i)}{d\omega}
    \right]
    \delta\left(\Delta E - \sum_{i=1}^n \omega_i\right)
    e^{- \int d\omega \frac{dI}{d\omega}}\, .
   \label{eq4.28}
\end{eqnarray}
These probabilities are referred to as quenching weights. In general, they have 
a discrete and a continuous part,\cite{Salgado:2002cd}
\begin{equation}
  P(\Delta E) = p_0\, \delta(\Delta E) + p(\Delta E)\, .
   \label{eq4.29}
\end{equation}
The discrete weight $p_0 = e^{- \int d\omega \frac{dI}{d\omega}}$ is a consequence of a finite
mean free path. It denotes the probability that 
no additional gluon is emitted due to in-medium scattering 
and hence no medium-induced energy loss occurs. 
Quenching weights have been calculated for several formalism of radiative parton energy loss.

If one treats the
medium-induced gluon energy distribution $\omega \frac{dI}{d\omega}$ 
explicitly as the medium modification of a
``vacuum'' distribution \cite{Salgado:2003gb}
\begin{equation}
 \omega \frac{dI^{({\rm tot})}}{d\omega} = 
 \omega \frac{dI^{({\rm vac})}}{d\omega} +
 \omega \frac{dI}{d\omega}\, ,
 \label{eq4.30}
\end{equation}
one can write the total probability of losing parton energy via vacuum or
medium-induced radiation as the convolution of (\ref{eq4.28}) with the probability
of vacuum energy loss
\begin{equation}
  P^{({\rm tot})}(\Delta E) = \int_0^\infty  d\bar{E} \, 
  P(\Delta E - \bar{E}) \, 
   P^{({\rm vac})}(\bar{E})\, . 
   \label{eq4.31}
\end{equation} 
Since gluon radiation in the vacuum underlies the 
scale dependence of fragmentation functions, this motivates models of medium-modified
fragmentation functions, in which the vacuum fragmentation function is convoluted with
(\ref{eq4.28}).
More specifically, if the parent parton loses with probability $P(\epsilon)$
an additional energy fraction $\epsilon = \frac{\Delta E}{E_q}$ 
prior to hadronization, then the leading hadron is a fragment of
a parton with lower energy $(1-\epsilon) E_q$; thus, it carries
a larger fraction $\frac{x}{1-\epsilon}$ of the initial parton energy. 
The inclusion of this effect amounts to replacing
the fragmentation function $D_{f\to h}(x,Q^2)$ in (\ref{eq2.1}) 
by the medium-modified
fragmentation function~\cite{Wang:1996yh,Gyulassy:2001nm}
\begin{eqnarray}
  D_{f\to h}^{(\rm med)}(x,Q^2) 
  = \int_0^1 d\epsilon\, P(\epsilon)\,
  \frac{1}{1-\epsilon}\, 
  D_{f\to h}(\frac{x}{1-\epsilon},Q^2)\, .
  \label{eq4.32}
\end{eqnarray}
For alternative approaches towards medium-modified fragmentation
functions, see Refs.\cite{Guo:2000nz,Osborne:2002dx,Armesto:2007dt,Domdey:2008gp}.

%%%%%%%%%%%%%%%%%%%%%%%%%%%%%%%%%%%%%%%%%
\subsubsection{BDMPS, Z, ASW, GLV, ... and all that}
\label{sec4.2.4}
A significant part of the literature on medium-induced gluon radiation can be related
to limiting cases of the gluon energy distribution (\ref{eq4.12}):
\begin{itemize}
	\item {\bf BDMPS}~\cite{Baier:1996kr,Baier:1996sk,Baier:2001yt} 
	is obtained from (\ref{eq4.12}) in the multiple soft scattering
	approximation (\ref{eq4.18}), using the limit
		\begin{equation}	
		  \omega \frac{dI^{\rm (BDMPS)}}{d\omega} (\omega_c\equiv\hat{q}L^2/2)
		  = {\rm lim}_{R=\omega_c L \to \infty}  \omega \frac{dI^{\rm (BDMPS)}}{d\omega} 
		  (\omega_c = {\rm fixed})\, .
		  \label{eq4.33}
		\end{equation}
	\item {\bf Z}akharov ~\cite{Zakharov:1996fv, Zakharov:1997uu,Zakharov:1998sv}
	has derived the BDMPS-result independently by a very
	different approach. In particular, he was the first to introduce the path-integral,
	with which equation (\ref{eq4.12}) can be written in a compact form. It would be
	historically correct to refer to this formalism as BDMPS-Z.
	\item {\bf ASW}~\cite{Wiedemann:2000za,Wiedemann:2000tf,Salgado:2003gb,Armesto:2003jh} 
	The specific form (\ref{eq4.12}) goes beyond \ref{eq4.33} in that it is
	valid for arbitrary values of $R=\omega_c\, L$. In this way, it includes effects from
	finite in-medium path length $L$ (which result e.g. in the infra-red cut-off of 
	medium-induced gluon radiation seen in Figure~\ref{fig9}) and it 
	accounts for the rescattering effects in the $k_T$-differential gluon emission. 
	This form was derived first in ~\cite{Wiedemann:2000za} and analyzed 
	for massless~\cite{Salgado:2003gb} and massive~\cite{Armesto:2003jh}.
	partons subsequently. It has been analyzed in the multiple soft scattering 
	approximation (\ref{eq4.18}) and in the $N=1$ opacity expansion.
	\item {\bf GLV} ~\cite{Gyulassy:2000er,Gyulassy:2001nm}
	While ASW compares $N=1$ opacity expansion and multiple soft 
	scattering limit, GLV focusses entirely on the opacity expansion of
	medium-induced gluon radiation, which had  been studied first in 
	 ~\cite{Wiedemann:2000za} and in ~\cite{Gyulassy:2000er}. To first
	 order in opacity, ASW and GLV obtain the same differential radiation
	 cross sections. 
\end{itemize}
We note that all above-mentioned formulations have been derived within the same
kinematic region
\begin{equation}
	E \gg \omega \gg \vert {\bf k}\vert \, ,\, 
	\vert {\bf q}\vert  \equiv \vert \sum_i {\bf q}_i \vert  \gg \Lambda_{\rm QCD}\, .
	\label{eq4.34}
\end{equation}
That means, the energy $E$ of the initial hard parton is much larger than the energy
of the emitted gluon, which is much larger than its transverse momentum ${\bf k}$ and
the transverse momentum ${\bf q}$ accumulated due to many scatterings of the projectile. 
By employing the above-mentioned formulations for phenomenological modeling, one 
inevitably extends their use beyond the range of their parametric validity (\ref{eq4.34}). 
In particular, to calculate a ${\bf k}_T$-integrated gluon energy distribution, one integrates
the transverse momentum over the entire kinematical range 
$\vert {\bf k}\vert \in \left[ 0, O(\omega)\right]$. Moreover, to account for large
parton energy loss, one allows for the case $\omega \sim O(E)$. Since exact
energy-momentum conservation at the vertex is lost with the approximations
(\ref{eq4.34}), numerical results of these integrals depend inevitably on the cut-off procedures.
For instance, the only difference between GLV and ASW lies in the implementation of these
cut-offs outside the region (\ref{eq4.34}). 
For the scope of the present review, we do not discuss these (important) details, which
are at the basis of an ongoing debate, and which are currently evaluated quantitatively
by the TECHQM Collaboration~\cite{TECHQM}. In the view of the present author, the best way 
to overcome the phenomenological limitations resulting from the approximation (\ref{eq4.34}) 
is to get rid of this approximation in the formulation of the medium-induced gluon
radiation. While this is a very challenging task in analytical formulations, it can be
achieved easily in Monte Carlo formulations, which we discuss in section~\ref{sec4.5}.

We note that there are at least two other formulations of medium-induced gluon 
radiation, which are currently used in phenomenological modeling of parton energy loss
\begin{itemize}
	\item {\bf Higher-twist formalism} \cite{Guo:2000nz,Wang:2001if,Wang:2001ifa}
	This is a calculation of medium-induced gluon
	radiation, which describes properties of the medium in terms of 4-point
	"higher-twist" matrix elements.  As for the formalisms discussed above, it can
	describe the interference between vacuum radiation and gluon radiation, which
	is phenomenologically important. This formalism shares many features with the
	above formalisms.
	\item {\bf AMY}~\cite{Arnold:2002ja} While all approaches discussed above involve modeling in 
	describing the interaction between projectile and target, this one does not. 
	It is the only dynamically consistent, model-independent formulation of 
	medium-induced gluon radiation, based solely on perturbative QCD. 
	The price to pay for this theoretical clean situation is the limitation to a peculiar
	kinematic region in which projectile energies are of the order of the temperature, 
	and where the temperature is very high ($T \gg T_c$) so that  
	hard-thermal-loop improved perturbation theory is
	applicable. Moreover, one neglects interference effects between vacuum 
	and medium-induced gluon radiation, which are known from other studies to affect
	numerical results by large numbers. 
	 As a consequence, for phenomenological
	applications one must extrapolate this formalism significantly outside its strict
	range of validity. 
\end{itemize}

%%%%%%%%%%%%%%%%%%%%%%%%%%%%%%%%
\subsection{Elastic interactions between projectile and medium}
\label{sec4.3}
Collisional mechanisms of parton energy loss, mediated via elastic interactions, have
been explored first in~\cite{Bjorken:1982tu,Thoma:1990fm,Braaten:1991we}.
It was then observed that at sufficiently high projectile energy, 
an essentially recoilless radiative energy loss mechanism is expected to dominate on
general kinematic grounds~\cite{Baier:1996sk,Zakharov:1997uu,Wiedemann:2000za,Gyulassy:2000er,Wang:2001if}. At sufficiently small projectile energies, 
however, recoil is expected to be non-negligible. Several recent model 
studies~\cite{Djordjevic:2006tw,Wicks:2005gt,Wicks:2007zz,Adil:2006ei,Peigne:2008nd}
attribute a sizable role to collisional mechanisms mediated via elastic interactions.

Elastic interactions between a partonic projectile $Q$ and degrees of freedom in the target
can transfer per unit path length a fraction $\Delta p_Q$ of the projectile momentum to
the target. Multiple interactions add incoherently for elastic processes, so that 
\begin{equation}
\frac{d\Delta p_Q}{dx}= \frac{1}{v_Q} \int dp_f (p-p_f) \int k^2 dk 
\left(n_q(k)\, \frac{d\sigma^{\rm int}_{Qq}(k,p_f)}{dp_f}\, +
 \frac{9}{4} n_g(k)\, 
\frac{d\sigma^{\rm int}_{Qg}(k,p_f)}{dp_f} \right)\, . 
\label{eq4.35}
\end{equation}
Here, $p$ is the initial and $p_f$ the final momentum of  the projectile $Q$, and 
$v_Q$ denotes its velocity in the rest frame of the medium. By $n_q(k)$ and $n_g(k)$,
we denote the distribution of quark and gluon scattering centers of momentum $k$
in the medium. The elastic scattering cross section can be written in the form
\begin{equation}
\frac{d\sigma^{\rm int}}{dp_f}=2\pi \int {\rm d}(\cos \psi) \frac{1}{4p^0 k^0} |{\cal{M}}|^2 d\Phi\, ,
\label{eq4.36}
\end{equation}
where $2\pi\int {\rm d}(\cos \psi)$ denotes the integration over the direction of the incoming
target particle, and $d\Phi$ denotes the phase space volume.

Within this framework, models for collisional energy loss calculations are fully specified 
in terms of the densities $n_q$, $n_g$ and the elastic scattering matrix element ${\cal M}$.
For the latter, one often uses the expression to lowest order in 
$\alpha_s$ with single gluon exchange in the t-chanel described by the HTL-resummed 
propagator. 
This is the starting point of many works on collisional energy loss, including the early 
works!\cite{Thoma:1990fm,Braaten:1991we}. 
There is a significant number of recent works, which implement variations of this
formalism, such as not including any assumption about the smallness of momentum
transfers~\cite{Djordjevic:2006tw}, calculating for constant coupling 
constant~\cite{Djordjevic:2006tw,Wicks:2005gt} or running coupling 
constant~\cite{Peshier:2006ah}.  Also, there are different models, which parametrize
the medium e.g. either as a set of massless particles with thermal momentum 
distribution~\cite{Djordjevic:2006tw,Wicks:2005gt}, 
or as a set of initially static massive scattering centers~\cite{Wicks:2007zz}. By making
the target scattering centers dynamical, these models parametrize not only the color
field strength but also the capacity of the medium to absorb recoil. 

We limit our discussion to some generic observations:
\begin{itemize}
         \item Elastic scattering cross sections are dominated by small-angle scattering
         involving small momentum transfer. As a consequence, the average collisional parton
         energy loss can be much larger than the typical parton energy loss encountered by
         a high-$p_T$-triggered particle, as discussed in the section~\ref{sec3.1}.
         	\item Some recent models result in an average collisional parton energy loss 
	which remains numerically significant over a wide transverse momentum 
	range~\cite{Djordjevic:2006tw,Wicks:2005gt,Peshier:2006ah}.
	Although the size of this effect can depend significantly on the modeling of the medium,
	this indicates that collisional effects must not be neglected in the description of 
	jet quenching. 
	\item While the mass-ordering (\ref{eq4.26}) of radiative energy loss depends on
	the projectile mass only, the mass-ordering of collisional energy loss is sensitive
	to the recoil properties of the medium and can be inverted compared to that
	of radiative energy loss~\cite{Kolevatov:2008bg}. 
\end{itemize}

%%%%%%%%%%%%%%%%%%%%%%%%%%%%%%%%%%%%%%%
\subsection{Monte Carlo Formulations of parton propagation in the medium}
\label{sec4.4}
In the absence of a medium, the final state parton showers, encoded in modern Monte 
Carlo (MC) event generators~\cite{Sjostrand:2006za,Corcella:2000bw,Gleisberg:2003xi},
can account reliably for jet event shapes, jet substructures and the main features of
other intra-jet characteristics, discussed in section~\ref{sec2.3}. To discuss jet quenching
on the level of multi-particle final states, it is of obvious interest to extend the applicability
of MC final state parton showers to medium effects. Any such attempt must address two central
issues:
\begin{itemize}
	\item {\it Specifying the spatio-temporal embedding of jets in matter.}\\
	In the absence of a medium, the length and time scales, over which branching processes 
	occur, do not enter the evolution.Hence, the final state parton shower is formulated 
	completely in momentum space. This is different in the presence of a medium, 
	when length scales and time scales determine which branching processes
	occur inside the medium and how frequently the medium can interact with the 
	parton shower. One may distinguish two aspects:\newline
	i) {\it Specifying the spatio-temporal evolution of the probe} amounts 
	to a somewhat model-dependent choice, since it is difficult to constrain
	 phenomenologically.  This choice should be made consistent with what is known 
	 parametrically	about the localization of partonic processes. In particular, a complete
	 spatio-temporal ordering can be specified by 
	attributing to each virtual parton in the branching process a lifetime of order
	$\tau_{\rm virtual \, life} = E/Q^2$ according to equation (\ref{eq2.4}). Once
	data for in-medium modified jets are compared to MC simulations, this
	picture of the spatio-temporal embedding of jets in matter can be scrutinized
	in an interplay between experiment and MC modeling.\newline
	ii) {\it The spatio-temporal extension and evolution of the medium} is important,
	since parton energy loss can depend strongly on in-medium path length. The
	current state of the art of modeling high-$p_T$ hadron suppression uses
	information from hydrodynamical simulations of heavy ion collisions, or simple
	parametrizations thereof. There is also a class of model studies, which specifies
	the geometrical distribution of matter from the nuclear overlap of Glauber 
	theory~\cite{Dainese:2004te,Eskola:2004cr}.
	\item {\it Specifying the interactions between projectile and medium}\\
	Simulations of the medium-modification of jets fragmentation can depend e.g. 
	on the strength and kinematics of interactions between partonic projectiles and
	the medium, on the relative weight of elastic and inelastic processes, and on the
	probability with which these interactions occur per unit path length. 
	In general, varying the nature of these interactions amounts to varying properties
	of the medium. Hence, the model-dependence which comes with this broad array 
	of conceivable interactions, is wanted. The extent to which the medium modification
	of jet fragmentation depends on assumptions about the interaction between projectile
	and medium determines the discriminatory power of jet quenching as a tool for
	characterizing properties of the medium. However, for the case of multiple
	interactions, the so-called non-abelian Landau-Pomeranchuk-Migdal effect,
	that is the destructive quantum interference between subsequent inelastic
	production processes, is known to strongly affect medium-induced radiative
	energy loss. Quantitative control of this interference is a prerequisite for
	determining the nature of interactions between projectile and medium.
\end{itemize}
In the following, we review shortly the current state of the art in encoding elastic and inelastic interactions with the medium in MC parton shower simulations. We focus mainly on formulations,
which aim at implementing what is known analytically about in-medium parton propagation,
as discussed in sections~\ref{sec4.2} and ~\ref{sec4.3}. For a self-contained presentation,
we start with some remarks about final state parton showers in the vacuum. 
In what follows, we do not discuss further
several Monte Carlo studies, which seem to account for the high-$p_T$ suppression patterns
at RHIC by invoking specific implementations of non-perturbative physics, see 
e.g.\cite{Zapp:2005kt,Cassing:2003sb,Werner:2008zza}. 
%
%%%%%%%%%%%%%%%%%%%%%%%%%%%%%%%%%%%%%%%%%%%%%
\subsubsection{Parton shower in the vacuum}
\label{sec4.4.1}
The MC algorithms for leading order final state parton showers in elementary collisions are 
documented in great detail in the 
literature~\cite{Sjostrand:2006za,Corcella:2000bw,Gleisberg:2003xi}. 
There are different variants, which differ e.g.
in the choice of evolution variable. Here, we recall but
some pertinent features, which are relevant for the subsequent discussion 
of medium effects. We do this by focussing on one particular MC implementation, the
mass-ordered parton shower used e.g. in the  \textsc{Pythia}~6.4 event
generator~\cite{Sjostrand:2006za}.
For each parton $a$ in a parton shower, the kinematics of its branching 
$a \to b+c$  is given in terms of the virtuality of the parent parton
and the momentum fraction $z$ of its total energy, which is carried by one 
of its daughters. The probability that
no splitting occurs between an initial and final virtuality $Q_i$ and $Q_f$,
respectively, is described by the Sudakov factor
\begin{eqnarray}
\label{eq4.37}
S_{\text a}(Q_{\rm i}^2,Q_{\text f}^2)
   =  \exp \left[ - \int \limits_{Q_{\text f}^2}^{Q_{\text i}^2} \!
     \frac{{\text d} Q'^2}{Q'^2} \!\!\int \limits_{z_-(Q'^2,E)}^{z_+(Q'^2,E)}
\!\!\!\!\!\!\!
     {\text d} z \,
     \frac{\alpha_{\text s}(z(1-z)Q'^2)}{2\pi}\, \sum_{\text b,c}
     \hat P_{{\text a}\to {\text bc}}(z) \right]\, . 
\end{eqnarray}
Here, $\hat P_{\text{a}\to\text{bc}}(z)$ are the standard LO parton splitting
functions for quarks and gluons ($a,b,c \in \{q,g\}$). The $z$-integral must be
regularized by an infrared cut-off scale below which parton splittings are
considered to be not resolvable. This defines $z_+$, $z_-$.
The probability density $\Sigma_{\text a}(Q_{\text i}^2,Q^2)$ for a splitting to 
occur at virtuality $Q^2$ is given in terms of the no-splitting probability 
(\ref{eq4.37}) as 
\begin{equation}
 \Sigma_{\text a}(Q_{\text i}^2,Q^2) 
      = \frac{{\text d} S_{\text a}(Q_{\text i}^2,Q^2)}{{\text d} (\ln Q^2)}
      = S_{\text a}(Q_{\text i}^2,Q^2) \sum_{\text b,c}
       \int \limits_{z_-(Q^2,E)}^{z_+(Q^2,E)}\!\!\!\!\!
     {\text d} z \, \frac{\alpha_{\text s}(z(1-z)Q^2)}{2\pi}
     \hat P_{{\text a}\to{\text bc}}(z)\, .
        \label{eq4.38}
\end{equation}
Here, the Sudakov form factor $S_{\text a}(Q_{\text i}^2,Q^2)$ denotes the 
probability for evolving from $Q_{\text i}^2$ to $Q^2$ without splitting, and 
the remaining factor is the differential probability for the splitting $a \to b+c$ at
$Q^2$, summed over all $b$ and $c$. 

To be specific: In this case, the parton shower for a parent parton $a$ of energy $E$ 
is initiated by determining its virtuality $Q_a^2$ according to the probability density 
$ \Sigma_{\text a}(E^2,Q^2)$. In accordance with (\ref{eq4.38}), one then selects
the type of parton splitting, and the momentum fractions of the daughters within
the kinematically allowed range $z\in \left[z_-(Q_a^2,E), z_+(Q_a^2,E)\right]$.
Then, one specifies the virtualities $Q_b$, $Q_c$ of the daughters subject to
some constraints: the virtuality should lie in between a hadronization scale and
the energy of the daugher, and the virtualities of the daughters should satisfy 
$Q_b^2+Q_c^2 < Q_a^2$. This branching is accepted if the momentum 
fraction $z$ chosen initially lie within the kinematically allowed range for daughter
masses $Q_b$, $Q_c$. Otherwise, new values for $Q_b$, $Q_c$ are chosen. 
If the branching is accepted, one reconstructs the full four-momentum for both 
daughters. Then, one treats them as seeds for subsequent branching processes. 
The partonic branching is terminated with the probability $S_{\text a}(Q_{\rm i}^2,Q_0^2)$
that no further branching occurs up to the hadronization scale $Q_0$. 

Implementing exact angular ordering in the parton shower is important to ensure
MLLA accuracy, see section ~\ref{sec2.3.3}. In the specific MC implementation
discussed above, this can be achieved by rejecting the generated $z$-values if
angular ordering is not satisfied. In other MC algorithms,  which use different 
ordering variables, this implementation of angular ordering is achieved differently. 
 
 From the above, is should be clear that any MC algorithm of jet quenching amounts 
 primarily to specifying how the no-splitting probability (\ref{eq4.37}) and the 
 probability density (\ref{eq4.38}) for branching can be supplemented by 
 medium effects. 
 
%%%%%%%%%%%%%%%%%%%%%%%%%%%%%%%%%%%%%%%% 
\subsubsection{Simulating elastic interactions}
\label{sec4.4.2}
Elastic interactions, in which the parton exchanges momentum and color with the
medium, can be modeled incoherently. In general, one can assume a density 
$n$ with which such elastic interactions occur along the path of the parton
projectile, and an elastic cross section $\sigma_{\rm el}$, which determines the
kinematics of the interaction. In close analogy to the Sudakov factor
(\ref{eq4.37}), one defines~\cite{Zapp:2008gi} then the probability that no scattering occurs 
within a time interval $\tau$
\begin{equation}
	S_{\rm no\,  scatt}(\tau) = \exp\left[  - \sigma_{\rm elas}\, n\, \tau \beta
\right]\, ,
	\label{eq4.39}
\end{equation}
where $\beta$ denotes the velocity of the parton and the density $n$ can be a 
function of the traveled distance $\beta\tau$.
If the time $\tau$ is specified to be the lifetime of a virtual parton prior to subsequent
branching, $\tau \sim E/Q^2$, then (\ref{eq4.39}) specifies the probability that this
subsequent branching occurs without prior interaction with the medium. 
With the probability $\left[1-S_{\rm no\,  scatt}(\tau)\right]$, the parton
scatters at some time $\tau'<\tau$. In this case, the parton exchanges momentum
with a scattering centre according to the differential elastic cross section
specified by the model. For instance, this could be the suitably regularized,
leading perturbative $t$-channel exchange term for quark-quark ($C_R = 4/9$), 
quark-gluon ($C_R = 1$) and gluon-gluon ($C_R = 9/4$) scattering
\begin{equation}
	\frac{{\rm d}\sigma_{\rm el}}{{\rm d}\vert t\vert} = \frac{\pi\alpha_s^2}{s^2} C_R
		\frac{s^2+u^2}{ |t|^2}\Bigg\vert_{\rm regularised}\, .
		\label{eq4.40}
\end{equation}
Multiple elastic scattering is implemented by specifying the momentum of the
outgoing partons from ${\rm d}\sigma_{\rm el}$ and further propagating them
according to (\ref{eq4.39}). 

This prescription interpolates between two controlled limiting cases: First, in 
the absence of a medium, one recovers the benchmark results of a vacuum 
parton shower. Second, in the case that many elastic interactions occur within
the lifetime of a virtual parton, vacuum branching processes become unimportant
and the code reproduces the results of analytic collisional energy loss calculations,
described in section~\ref{sec4.3}. In addition, this MC technique provides easy 
access to information 
about the dynamical fluctuations around the event average, and to information about \
multi-particle final states, such as distributions of recoil partilces.

%%%%%%%%%%%%%%%%%%%%%%%%%%%%%%%%%%%%%%%%%%%%% 
\subsubsection{Simulating inelastic processes}
\label{sec4.4.3}
MC implementations of medium-induced parton energy loss started in the early 90's
 with \textsc{Hijing}~\cite{Wang:1991hta}, which simulates complete nucleus-nucleus 
 collisions and includes a simplified model for radiative energy loss. More recently, 
there have been several developments:  there
 is  \textsc{Pyquen}~\cite{Lokhtin:2005px}, which modifies {\it a posteriori} the
standard \textsc{Pythia~6.2} jet events essentially by reducing the
energy of partons in the shower and by adding additional gluons to that
shower according to distributions motivated by parton energy loss calculations. 
The first version of the final state parton shower \textsc{JEWEL}~\cite{Zapp:2008gi} 
implements elastic energy loss as described in section~\ref{sec4.4.2} and it
mimics radiative energy loss by multiplying the vacuum splitting function in the
parton shower with a factor $\left( 1+ f_{\rm med}\right)$~\cite{Borghini:2005em}. 
A future version of  \textsc{JEWEL}
will simulate medium-induced radiation by an algorithm based on formation time,
which is known to account for non-abelian quantum interference as described
in the context of equation (\ref{eq4.42}) below~\cite{Zapp:2008af}. 
\textsc{Q-PYTHIA}~~\cite{Armesto:2008qh} simulates medium-modifications of a 
final state parton shower
by modifying the splitting functions in a PYTHIA vacuum shower as described below.
There is also a model  \textsc{YaJEM} (Yet Another Jet Energy-loss Model),
where the virtuality of partons in the shower is increased due to 
interactions with the medium thus stimulating additional gluon emissions~\cite{Renk:2008pp}.
This implements a conceivable microscopic dynamics which is not yet
supported by analytical calculations. There is a recent comparison study of
several medium-modified QCD evolution shower scenarios~\cite{Renk:2009nz}.

 In this section, we focus mainly on MC algorithms which
 \begin{enumerate}
	\item aim at simulating the medium-modification of the entire parton fragmentation
	pattern rather than focussing on the energy degradation of the most
	energetic partons only. In particular, the MC algorithm should reduce to a 
	phenomenologically viable vacuum parton shower in the
	absence of medium effects.
	\item aim at a dynamical MC implementation of the analytical results of
	medium-induced gluon radiation reviewed in section~\ref{sec4.2}. 
 \end{enumerate}
 Two different methods of simulating medium-modified parton showers
aim at satisfying these criteria:
 
\noindent
{\bf The method of medium-modified splitting functions}
Inelastic interactions between projectile and the medium may be regarded as a source of an
additional medium-induced branching $a \to b+c$ of the projectile parton $a$. For instance,
within the BDMPS formalism for parton energy loss, 
medium-induced gluon radiation can be written formally as a modification of the vacuum
splitting function, see e.g. equation (\ref{eq4.30}) and subsequent discussion. This
motivates an approach, in which the splitting function $\hat{P}_{a\to bc}(z)$ 
 in the vacuum parton shower defined by (\ref{eq4.35}), (\ref{eq4.36}) is replaced 
 by a medium-modified splitting function
 \begin{equation}
 	\hat{P}_{a\to bc}(z)^{\rm tot} = \hat{P}_{a\to bc}(z)^{\rm vac} +
		\Delta \hat{P}_{a\to bc}(z,Q^2,\hat{q},L,E)\, .
		\label{eq4.41}
 \end{equation}
The MC model \textsc{Q-Pythia}~\cite{Armesto:2008qh} follows this approach and it
specifies the medium-modification \newline
$\Delta \hat{P}_{a\to bc}(z,Q^2,\hat{q},L,E)$ of the 
parton branching in terms of the analytically known gluon energy distribution 
$\omega dI/d\omega$. As such, the medium-modification in (\ref{eq4.41}) does not
depend solely on the momentum fraction $z$, but it depends also on the in-medium
path length $L$, the energy of the projectile $E$, properties of the medium which we
characterize here by the quenching parameter $\hat{q}$ and the virtuality $Q^2$
of the projectile quark. The current version of  \textsc{Q-Pythia} aims at taking 
coherence effects into account by attributing the coherence time (\ref{eq2.5}) to
the radiated gluon and calculating subsequent medium-induced gluon emissions 
for a reduced in-medium path length $L-l_{\rm coh}$, where $l_{\rm coh} = 2\omega/k_T^2$.
%
%%%%%%%%%%%%%%%%%%%%%%%%%%%%%%%%%%%%%%%%%%%%%%%%%%%%%%%%%%%%%%%%%%%%
\begin{figure}[h]\epsfxsize=12.7cm
\centerline{\epsfbox{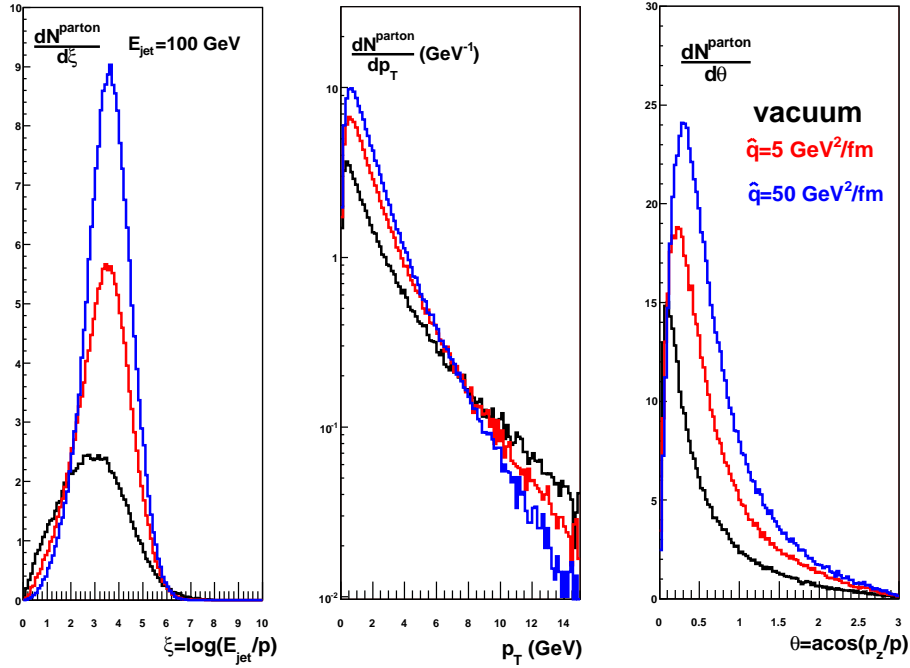}}
\caption{
Results of the \textsc{Q-Pythia} MC final state parton
shower for the intrajet parton distributions in $\xi = \ln\left( E_{\rm jet}/p\right)$ (left),
$p_T$ (middle) and $\theta = \arccos (p_z/p)$ (right) of a gluon jet with energy
$E_{\rm jet} = 100$ GeV. The vacuum baseline is compared to a medium of length
$L=2$ with transport coefficient $\hat{q}=5$ and 50 ${\rm GeV}^2/{\rm fm}$.
Figure taken from Ref.~\cite{Armesto:2008qh}.
}
\label{figqphythia}
\end{figure}
%%%%%%%%%%%%%%%%%%%%%%%%%%%%%%%%%%%%%%
%
%
In this way, \textsc{Q-Pythia} is automatically consistent with the vacuum baseline in the
absence of a medium, and it reproduces the analytically known results for the 
medium-modification of the single gluon energy distribution. Moreover, this proposal
has the advantage that the replacement (\ref{eq4.41}) is relatively easy to implement
in existing event generators. 
On the other hand, the prescription (\ref{eq4.41}) does not
allow to follow dynamically the recoil from the projectile parton to the medium - this
may be important e.g. for the description of the soft particle yield associated to highly
energetic jets. A more fundamental problem can be understood by recalling 
that a typical parton shower will also include gluons, which are fully formed in 
the sense of independent quanta, well before the partonic projectile has traversed
the entire medium. The splitting function of such a gluon should not 
depend on the maximal remaining in-medium path length $L-l_{\rm coh}$, 
simply since the gluon production process is completed before the gluon "sees"
this length. This shows that  \textsc{Q-Pythia} remains, strictly speaking, a heuristic
proposal, although it encodes much of what is known
analytically about radiative parton energy loss.

\noindent
{\bf Probabilistic implementation of non-abelian quantum interference}
If the time between subsequent interactions of a partonic projectile and the medium
is much larger than the formation time (\ref{eq2.5}) of the gluon produced in an
inelastic process, then this inelastic process is an incoherent gluon production,
which can be iterated probabilistically. The challenge for any probabilistic MC
implementation of the medium-dependence of inelastic processes is the opposite case
that the gluon formation process extends over more than one interaction between
projectile and medium. Then, spatially separated interactions act coherently
and quantum interference becomes important for gluon production.
It has been demonstrated ~\cite{Zapp:2008af} that this quantum interference can be implemented in a 
probabilistic MC algorithm by requiring that the momentum transfer from different 
scattering centers to the partonic projectile acts  totally coherently for gluon production, 
if it occurs within the formation time $t_f$, and that it acts incoherently, if it occurs outside $t_f$.

The proposed MC algorithm is as follows~\cite{Zapp:2008af}:
To be specific, consider a medium,  characterized by a distribution of target scattering 
centers $Q_T$,  which present inelastic $\sigma^{qQ_T \to qQ_T g}$ and elastic
$\sigma^{qQ_T \to qQ_T}$ cross sections to a projectile $q$. Consider the case 
that the MC algorithm specifies that a particular scattering center $Q_T$ is involved
in an inelastic process. If the initial formation time 
\begin{equation}
t_f = 2\omega/{\bf k_{{\rm init}}}^2
\label{eq4.42}
\end{equation} 
of  that gluon is shorter than the
distance to the next scattering center, then the gluon is formed in this 
interaction. The outgoing gluon and quark are propagated further as independent
degrees of freedom. In the opposite case, the projectile components emerging from
the first interaction will interact a second time before the gluon is formed. At this
second interaction, production of an additional gluon is not possible, since it
interferes destructively with the not yet fully formed gluon. So, the second scattering 
center is the source of an elastic interaction of  transverse 
momentum $q_T$ transfered to the projectile components. The MC algorithm 
adds $q_T$ to the transverse momentum of the first interaction and it
recalculates the inelastic gluon 
production under the assumption that both scattering centers act
effectively as one coherent scattering center. In practice, this increases 
the transverse momentum between the outgoing quark and gluon to
${\bf k_{{\rm init}}} \to {\bf k_{{\rm init}}} + {\bf q}$, and hence it shortens the 
formation time to 
\begin{equation}
t'_f= 2\omega / \left(  {\bf k_{{\rm init}}} + {\bf q}  \right)^2\, .
\label{eq4.43}
\end{equation}
If this shortened formation time $t'_f$ is shorter than the distance to the next (third)
scattering, then the gluon is regarded as fully formed. Otherwise, the procedure is
repeated and the elastic momentum transfer of this next scattering center is added 
to $k_{\perp{\rm init}} + q_T$.

The above procedure provides a probabilistic implementation of (\ref{eq4.12}).
This can be seen very clearly on the level of the $N=1$ opacity expansion: 
Expressing the initial transverse gluon momentum ${\bf k_{\rm init}}$ in terms
of the outgoing transverse gluon momentum ${\bf k_{\rm init}} = {\bf k}-{\bf q}$,
one finds that the gluon formation times $t_f$, $t'_f$ prior to and after secondary
scattering correspond exactly to the formation times $\tau_1$, $\tau$ in the
$N=1$ opacity result (\ref{eq4.15}). In particular, the interference term
$\left[ L/\tau_1 - \sin L/\tau_1 \right]$ in (\ref{eq4.15}) prevents gluons from being
radiated prior to interacting at distance $L$, if their formation time is too long.
This is exactly the coherence condition in the MC algorithm described above.

The algorithm described here is consistent with what is known qualitatively and 
quantitatively about  medium-induced gluon radiation (\ref{eq4.12}). In particular,
it reproduces the $L^2$-dependence of the average energy loss (\ref{eq4.25})
and the characteristic $\sqrt{\omega}$-modification of the gluon energy 
distribution in the limit of multiple soft scatterings. The present author expects
that the same formation time physics can account reliably for multi-gluon emission.
However, very little is known about the medium-modification of multi-gluon emissions
and, strictly speaking, the algorithm remains at present a credible but heuristic 
extrapolation to this case.

%%%%%%%%%%%%%%%%%%%%%%%%%%%%%%%%%%%%%%%%%%%%
\subsection{Applying the AdS/CFT correspondence to in-medium parton propagation}
\label{sec4.5}
The evolution of a parton shower is a weak coupling phenomenon
as long as the virtuality of partons lies well above the hadronic scale.
In fact, irrespective of whether parton splitting occurs in the vacuum
or in the medium, one expects the coupling at the splitting vertex to be perturbatively
small if the virtuality of the parent parton is sufficiently high. 
However, the interaction between components of the parton shower and
the medium may involve multiple soft momentum exchanges, which may occur
with non-perturbatively large coupling. This motivates the application of strong coupling
techniques to in-medium parton propagation. 

Finite temperature lattice QCD has been the main calculational tool for first principle
calculations of medium properties at non-perturbatively large coupling~\cite{Laermann:2003cv}. However, information on real-time dynamics in a strongly coupled quark gluon plasma is difficult
to extract from lattice QCD. For a large class of non-abelian thermal gauge theories,
the Anti-de Sitter / Conformal Field Theory (AdS/CFT) conjecture offers an alternative. 
The AdS/CFT conjecture asserts that in the limit of large number of colors ($N_c$) and 
large t'Hooft coupling $\lambda = g^2\, N_c$, a certain class
of non-abelian gauge field theories is equivalent to the supergravity limit of a
corresponding dual string theory, with the background metric describing a curved
five-dimensional anti-deSitter space, see Ref.~\cite{Aharony:1999ti} for a review
of the original literature. This AdS/CFT correspondence 
also applies to the thermal sector, if the five-dimensional
anti-deSitter space is endowed with a Schwarzschild black hole. In practice,
the AdS/CFT correspondence provides a technique for doing complicated
or otherwise intractable calculations in the strong coupling limit of a quantum
field theory by solving a comparatively simpler semi-classical problem in the
dual gravity theory. 

The gravity dual of QCD is not known. One may identify two main motivations for 
applying the AdS/CFT correspondence to problems in heavy ion physics. First, on a
qualitative level, there are open conceptual issues in QCD for which one may 
want to seek guidance in other non-abelian field theories for which strong coupling
techniques exist. For instance, the strongly coupled plasmas described by AdS/CFT
are known to be free of quasi-particles, and the picture of the QCD plasma
as a system of weakly coupled quasi-particles is on unclear footing, but most 
derivations of QCD transport and hydrodynamics rest on it. Here, the AdS/CFT 
correspondence can shed light on the conceptual issue, which features of 
transport theory persist beyond
the quasi-particle picture. Second, on a quantitative level, applications of the 
AdS/CFT correspondence to QCD must ultimately involve an argument 
that what has been calculated for a large class of non-abelian quantum field theories 
with gravity dual (which share many features with QCD), can be used as guidance for QCD. 
This assumption is supported by several observations. For instance, the energy
density and pressure, calculated with the help of the AdS/CFT correspondence,
is three quarters of the Stefan-Boltzmann value in the limit of infinitely strong 
coupling~\cite{Gubser:1996de} and it is slightly larger for large but finite $\lambda$,
in good agreement with the value found in QCD lattice calculations for temperatures above
but at the order of the critical temperature $T_c$. Another well-known example
is the calculation of the ratio $\eta/s$ of shear viscosity over entropy 
density~\cite{Policastro:2001yc}. This 
quantity turns out to take the same universal and extremely 
low value $1/4\pi$ in the strong coupling limit of all quantum field theories with a 
gravity dual. On the one hand, the universality of this result may be taken as an
indication that the value of $\eta/s$ in a QCD plasma is also very low. 
This assumption is supported by first exploratory lattice calculations~\cite{Meyer:2007ic}, 
and it is in line with hydrodynamical simulations of RHIC data. 
These and other examples have prompted in recent years applications of the
AdS/CFT correspondence to the calculation of many quantities, which are of
phenomenological interest for heavy ion physics. Here, we focus solely on 
two conceptually different approaches to studying parton energy loss
with the help of the AdS/CFT correspondence:

\noindent
{\bf Calculation of the quenching parameter}~\cite{Liu:2006ug}
This approach does not aim at applying strong coupling techniques to all aspects of 
in-medium parton propagation. Rather, it takes into account that at sufficiently
high virtuality, parton branching in QCD is best described perturbatively. 
In the multiple soft scattering limit, the only non-perturbative input to the medium-induced
gluon energy distribution (\ref{eq4.12}) is the quenching parameter $\hat{q}$,
which is defined as the short distance behavior of a light-like Wilson loop,
\begin{equation}
	\langle W^A \left( {\cal C}_{\rm light-like}\right)\rangle 
	= \exp \left[ - \frac{1}{4\sqrt{2}} \hat{q}\, L^-\, r^2\right]\, .
	\label{eq4.44}
\end{equation}
Here, ${\cal C}_{\rm light-like}$ denotes a closed path, consisting of two very 
long parallel pieces of length $L^-$ along the light cone, which are separated
by the transverse distance $r$. As explained in detail in Ref.\cite{Liu:2006he}, 
this definition follows from the observation that in the high energy limit, the path 
integral (\ref{eq4.11}) reduces to (\ref{eq4.44}). For the ${\cal N}=4$ 
super-symmetric Yang Mills (SYM) theory, the result for large t'Hooft coupling 
$\lambda = g^2\, N_c$ has been calculated by use of the AdS/CFT 
correspondence~\cite{Liu:2006he}
\begin{equation}
\hat{q}_{SYM} = \frac{\pi^{3/2}\Gamma(3/4)}{\Gamma(5/4} \sqrt{\lambda} T^3
\approx 26.69\, \sqrt{\alpha_{\rm SYM}\, N_c} \, T^3\, .
\label{eq4.45}
\end{equation}
If one would relate this to QCD by fixing $N_c =3$ and $\alpha_{SYM} = .5$, then
$\hat{q}_{SYM} = 4.4$ GeV$^2$/fm for a temperature of $T = 300$ MeV. This number
is comparable to the values extracted in comparisons of jet quenching models to 
RHIC data. For different theories with gravity dual, one finds
that the quenching parameter scales with the square root of the entropy density~\cite{Liu:2006ug}. 
If QCD follows this behavior, then $\hat{q}_{QCD}/\hat{q}_{SYM} = 
\sqrt{47.5/120} \approx 0.63$. Moreover,
\begin{equation}
	\hat{q}_{\rm LHC} = \sqrt{\frac{dN_{\rm ch}^{\rm LHC}/d\eta}{dN_{\rm ch}^{\rm RHIC}/d\eta}}\, 
		\hat{q}_{\rm RHIC}\, .
		\label{eq4.46}
\end{equation}
Other calculations of the quenching parameter $\hat{q}$, and their relation to 
observations at RHIC are discussed in Ref.~\cite{Liu:2006he}.

\noindent
{\bf Drag calculations of parton energy loss}~\cite{Herzog:2006gh,Gubser:2006bz,CasalderreySolana:2006rq}
This approach applies the AdS/CFT strong coupling technique to all aspects of in-medium 
parton energy loss. One considers a heavy quark of velocity $v$ and one calculates the
force needed to maintain this velocity, that is, the force needed to drag the quark through
the medium at velocity $v$. This allows one to determine the medium-induced
energy loss of the quark at velocity $v$,
\begin{equation}
	\frac{dE}{dx} = - \frac{\pi}{2}\, \sqrt{\lambda}\, T^2\, \frac{v}{\sqrt{1-v^2}}\, .
	\label{eq4.47}
\end{equation}
We remark that the derivation of (\ref{eq4.47}) is valid for sufficiently low velocities,
\begin{equation}	
	\sqrt{\gamma} < \frac{M}{\sqrt{\lambda}\, T}\, ,
	\label{eq4.48}
\end{equation}
while the derivation of the quenching parameter is valid only in the opposite limit
of sufficiently small quark masses $M$ or sufficiently high parton velocities, when
$\sqrt{\gamma} > \frac{M}{\sqrt{\lambda}\, T}$. This is one of the reason which has
made the comparison of drag calculations and calculations of the quenching 
parameter difficult. Arguments have been put forward that for velocities exceeding
(\ref{eq4.48}), the drag result (\ref{eq4.47}) will be small compared to energy loss due
to acceleration~\cite{Kharzeev:2008he,Fadafan:2008bq}. In a particular toy model,
energy loss due to radiation of an accelerated quark was shown to interfer destructively
with energy loss due to drag~\cite{Fadafan:2008bq}.
The velocity dependence of the drag result (\ref{eq4.47}) is characteristically different
from that of QCD. However, based on a heuristic interpretation of (\ref{eq4.47}) in terms of
a quasi-particle picture, one has argued for a close analogy of the drag result and
the QCD result for radiative parton energy loss~\cite{Dominguez:2008vd}.

In contrast to calculations of the quenching parameter, 
drag calculations of parton energy loss provide a complete dynamical description
of parton energy loss, assuming the same non-perturbatively large coupling constant
to all stages of the problem. While this treatment neglects the scale-dependence of the
QCD coupling constant, it has the advantage that the interaction between partonic
projectile and medium can be formulated without any further model assumption.
Dragged quarks are well-defined sources of hydrodynamic perturbations in
non-abelian plasmas. In particular, one finds that at least some fraction of the
energy, deposited by a dragged quark in the medium, propagates via sound waves
and gives rise to a Mach cone~\cite{Gubser:2007ga}. A lot of work on 
the radiation pattern associated with
dragged quarks has been motivated by suggestions that this Mach cone like structure
survives hadronization and that it may have been observed already at RHIC in the 
angular dependence of high-$p_T$ triggered two-particle correlations.

%%%%%%%%%%%%%%%%%%%%%%%%%%%%%%%%%%%%%%%%%%%%%

\end{document}